\documentclass[prb, aps, showpacs, floatfix, notitlepage, reprint, nofootinbib, onecolumn, 12pt]{revtex4-2}
\usepackage{graphicx}
\usepackage[ansinew]{inputenc}
\usepackage{array}
\usepackage{color}

\usepackage{amsmath}
\usepackage{amsxtra}
\usepackage{amstext}
\usepackage{amssymb}

\usepackage{latexsym}
\usepackage{dsfont}
\usepackage{lipsum}
\usepackage[colorlinks=true,allcolors=blue]{hyperref}

\usepackage{enumitem}
\setenumerate[1]{label=(\Roman*)}

\usepackage[left=25mm,right=25mm,bottom=20mm,top=20mm]{geometry}

\renewcommand{\Re}{\text{Re}}
\renewcommand{\Im}{\text{Im}}

\newcommand{\mb}[1]{\mathbf{#1}}
\newcommand{\bs}[1]{\boldsymbol{#1}}
\newcommand{\tm}[1]{\text{#1}}
\newcommand{\lel}{\left}
\newcommand{\rer}{\right}

\newcommand{\avg}[1]{\left\langle #1\right\rangle}

\newcommand{\avgg}[1]{\left\langle\left\langle #1\right\rangle\right\rangle}
\newcommand{\bra}[1]{\left\langle{#1}\right|}
\newcommand{\ket}[1]{\left|{#1}\right\rangle}
\newcommand{\comm}[2]{\left[#1,#2\right]}

\newcommand{\bj}{\mathbf{j}}
\newcommand{\bk}{\mathbf{k}}
\newcommand{\bL}{\mathbf{L}}
\newcommand{\bp}{\mathbf{p}}
\newcommand{\bP}{\mb{P}}
\newcommand{\cP}{\mathcal{P}}
\newcommand{\br}{\mathbf{r}}
\newcommand{\bral}{\mathbf{r}_\alpha}

\newcommand{\dbral}{\dot{\mathbf{r}}_\alpha}

\newcommand{\ddbral}{\ddot{\mathbf{r}}_\alpha}

\newcommand{\bR}{\mathbf{R}}

\newcommand{\bx}{\mathbf{x}}
\newcommand{\cdd}{\cdot}

\newcommand{\om}{\omega}

\newcommand{\al}{\alpha}

\newcommand{\dd}[2]{\frac{d #1}{d{#2}}}
\newcommand{\pd}[2]{\frac{\partial #1}{\partial{#2}}}

\newcommand{\curl}[1]{\boldsymbol{\nabla}_{#1}\times}

\newcommand{\dive}[1]{\boldsymbol{\nabla}_{#1}\cdot}
\newcommand{\grad}[1]{\boldsymbol{\nabla}_{#1}}

\def\XXint#1#2#3{{\setbox0=\hbox{$#1{#2#3}{\int}$}
     \vcenter{\hbox{$#2#3$}}\kern-.5\wd0}}

\newcommand{\intw}{\int_{-\infty}^{\infty} d\om\;}
\newcommand{\intwd}{\int_{-\infty}^{\infty} \frac{d\om}{2\pi}\;}
\newcommand{\intwp}{\int_{-\infty}^{\infty} d\om'\;}

\newcommand{\intwhp}{\int_{0}^{\infty} d\om'\;}

\newcommand{\intx}{\int d^3x \;}

\newcommand{\intr}{\int d^3r \;}

\newcommand{\intR}{\int d^3R \;}

\newcommand{\intk}{\int d^3k \;}

\newcommand{\Ah}{\mathbf{A}}

\newcommand{\Eh}{\mathbf{E}}

\newcommand{\Hh}{\mathbf{H}}
\newcommand{\Bh}{\mathbf{B}}
\newcommand{\Lh}{\mathbf{L}}

\newcommand{\Ph}{\mathbf{P}}

\newcommand{\Rh}{\mathbf{R}}
\newcommand{\Mh}{\mathbf{M}}

\newcommand{\jh}{\mathbf{j}}
\newcommand{\mh}{\mathbf{m}}
\newcommand{\rh}{\mathbf{r}}
\newcommand{\ph}{\mathbf{p}}

\newcommand{\thh}{\boldsymbol{\theta}}

\newcommand{\Lhh}{\mathbf{\widehat{L}}}

\newcommand{\Phh}{\mathbf{\widehat{P}}}
\newcommand{\Ohh}{\mathbf{\widehat{O}}}
\newcommand{\Rhh}{\mathbf{\widehat{R}}}
\newcommand{\Mhh}{\mathbf{\widehat{M}}}

\newcommand{\mhh}{\mathbf{\hat{m}}}

\newcommand{\jhh}{\mathbf{\hat{j}}}
\newcommand{\rhh}{\mathbf{\hat{r}}}
\newcommand{\phh}{\mathbf{\hat{p}}}

\begin{document}
\title{Is there a conflict between causality and diamagnetism?}
%\title{On the resolution of the causality-diamagnetism conflict}
\author{Niclas Westerberg}
\email{Niclas.Westerberg@glasgow.ac.uk}
\author{Stephen M. Barnett}
\affiliation{School of Physics and Astronomy, University of Glasgow, Glasgow, G12 8QQ, United Kingdom}

\begin{abstract}
There is a long-standing apparent conflict between the existence of diamagnetism and causality as expressed through the Kramers-Kronig relations. In essence, using causality arguments, along with a small number of seemingly well-justified assumptions, one can show that diamagnetism is impossible. However, experiments show diamagnetic responses from magnetic media. We present a resolution to this issue, which also explains the absence of observed dia-electric responses in media. In the process, we expose some of the short-comings in earlier analyses that have kept the paradox alive.
\end{abstract}

\date{\today}
\maketitle

\section{Introduction}
A diamagnet opposes applied magnetic fields. This is, of course, well known and has been measured and treated theoretically countless times in the past century, found in for instance Refs.~\cite{jackson, landau, stratton} as well as Ref.~\cite{NoltingRamakanth}. The effect is typically weak, and linear in the applied magnetic field, given a weak enough applied field.\footnote{Here we mean that the magnetic interaction energy must be smaller than the other energies of the system.} With the exception of superconductors, it can only be measured in the absence of all other magnetic responses. Diamagnetism is therefore only observable in some materials. We will refer to these as diamagnetic materials, or diamagnets, of which water and bismuth are common examples \cite{handbook}. However, it is ubiquitous and present in some form or another in all materials \cite{vanVleck}, including single atoms \cite{rydbergAtoms}. Nevertheless, it is also known that there is a long-standing problem with our treatment and understanding of diamagnetism: it seems to be in conflict with the principle of causality. To be precise, one can show that diamagnetism ought to be impossible, based on the seemingly innocuous assumptions that the diamagnet (1) obeys causality and (2) is passive (i.e., energy is lost, rather than gained inside a diamagnetic medium) \cite{landau}. These assumptions, as embodied in the Kramers-Kronig relations \cite{kramers, kronig, landau}, produce the inconsistency. As such, this has remained unsolved for nearly a century. 

Nevertheless, diamagnets clearly do exist \cite{handbook}. Yet, this theoretical inconsistency does mean that we cannot work with diamagnetic materials in macroscopic quantum electrodynamics \cite{scheel}. Over the years, proposed solutions have included everything from questioning the meaning of magnetisation at high frequencies \cite{landau, ModifiedKKSilveirinha2011}, to the involvement of spatial dispersion \cite{pitaevskii2012, alusPaper, norway_diamagnetism_1, norway_modifiedKK}, attempts to circumvent the passivity assumption \cite{markel}, and the stipulation that diamagnetism is a nonlinear effect and therefore precluded from linear response theory \cite{scheel, scheelUnified}.\footnote{The authors note that, since the history of this issue is long, we have decided to collate the attempts into themes rather than mentioning each and every one.} Most commonly, however, the issue is simply ignored, or the theory is artificially applied away from diamagnetic materials. 

Addressing this theoretical inconsistency between causality and diamagnetism is becoming increasingly important, as metamaterials have made the explicit manipulation of magnetic responses possible \cite{pendryDiamagnetism, soukoulis2008strongDiamagnetism, planarMetamaterialMagnetism, layeredDiamagnetism, layeredDiamagnetism2, temporalDiamagnetism, chiralMetadiamagnetism}. An effort has thus been made to find ways to incorporate the magnetic response in the metamaterial modelling. 
A variety of methods have been employed in this goal, from calculating local-field effects in split-ring resonators \cite{soukoulis2008strongDiamagnetism} and computing the effective long-wavelength response in layered materials \cite{planarMetamaterialMagnetism, layeredDiamagnetism, layeredDiamagnetism2} to chiral dielectric meta-atoms \cite{chiralMetadiamagnetism} and temporal modifications \cite{temporalDiamagnetism}. Diamagnetic responses have been found in this way, but the techniques vary and are not easily applicable outside the realm of metamaterials: they rely on, for instance, spatial dispersion within the metamaterial unit cell and modifications to the notion of the multipolar expansion \cite{alusPaper} or numerical simulations of the current response \cite{belov2013broadband}. In this work we take a broader approach and offer a different perspective, aiming to explain the phenomenon as a whole rather than specific scenarios.

We provide a resolution to this diamagnetism-causality conundrum that applies equally well to ordinary matter as to metamaterials, and discuss how previous attempts fit into the new picture. We focus on insulators, but expect similar arguments to be possible for conductors when the applied field is weak.

To set the context, we will concern ourselves with magnetism in linear macroscopic media, where in general, we can connect the medium magnetisation $\mb{M}$ to the magnetic induction field $\mb{B}$ through the susceptibility $\chi$, which is defined through
\begin{align}\label{eq:chiDef}
\mu_0 \Mh(\om) = \chi(\om) \Bh(\om),   
\end{align}
where $\mu_0$ is the permeability of free space. We note that this relationship is often written in terms of the magnetic field, $\Hh$. In our approach it is more natural to work with the magnetic induction, but working with $\Hh$ leads to the same conclusion. For simplicity, we restrict out analysis to isotropic media but anticipate no difficulties in treating more general media characterised by a susceptibility tensor. We here define diamagnetism as a statement about the magnetostatic response, i.e. $\chi(\om\rightarrow 0)$, as this is where the issue lies. In particular, we call a medium diamagnetic if
\begin{align}
\chi(0) < 0,    
\end{align}
meaning that the medium magnetisation opposes the applied magnetostatic field. 

This manuscript is structured as follows: in Section~\ref{sec:Problem}, we discuss and formulate the assumptions that lead to the apparent conclusion of diamagnetism being impossible. We follow this by Section~\ref{sec:PrevArguments} where we discuss the previous attempts at reconciling theory with the observations of diamagnetism and the limitations of these. In Section~\ref{sec:Resolution}, we present the resolution to the diamagnetism-causality conflict, the implications of which we discuss in Section~\ref{sec:Conclusion}.

\section{Reasonable assumptions and a contradiction.}\label{sec:Problem}
Let us start by convincing ourselves that there is indeed a conflict between diamagnetism and causality. We note that this can be equivalently formulated in terms of the magnetic field $\Hh = \Bh/\mu_0-\Mh$, where we instead write $\Mh(\om)=\chi^H(\om)\Hh$. The two susceptibilities are related by $\chi^H(\om) = \chi(\om)/[1-\chi(\om)]$, which does not change the analytical, or causal, properties. We have chosen to work with $\mu_0\Mh(\om) = \chi(\om)\Bh(\om)$ as the magnetic induction is more directly connected with the vector potential $\Ah$ defined such that $\Bh = \curl{}\Ah$. With this in mind, the argument is simple and based on three assumptions: 
\begin{enumerate}
    \item \label{as:causality} Causality: the medium response function is causal, as in, the magnetisation cannot depend on future times. The mathematical expression of this is
    \begin{align}
    \mu_0\mb{M}(t) = \int_{-\infty}^{t}dt'\; \chi(t-t')\mb{B}(t').    
    \end{align} 
    If we use the Fourier transform convention $f(t) = (2\pi)^{-1/2}\int d\om \; e^{-i\om t} f(\om)$, then this manifests itself as demanding that $\chi(\om)$ is an analytic function on the upper half of the complex $\om$-plane.
    
    \item \label{as:reality} We require that the magnetisation $\mb{M}$ as a function of time is a real function. This is because, like the magnetic induction $\mb{B}$, it is a measurable quantity. In frequency-space, this implies the oft-used reflection property \cite{jackson}: $\chi(-\om) = \chi^*(\om^*)$.
    
    \item \label{as:passivity} The medium is passive, or in other words, energy cannot be gained as a function of time. This usually is expressed through the assumption that $\Im\,\chi(\om) \geq 0$ at all frequencies $\om$.
\end{enumerate}
It is remarkable that assumptions \ref{as:causality}-\ref{as:passivity}, taken together, expose a direct conflict with the existence of diamagnetism. This is a long-standing problem in macroscopic magnetism \cite{landau}, but we repeat the arguments here. Interestingly, causality links dispersion and absorption and as a consequence, the magnetostatic $\omega\rightarrow 0$ response is connected to the high-frequency behaviour of the susceptibility. This is embodied in the Kramers-Kronig relations \cite{kramers, kronig}. Therefore, given that $\chi(\om)$ decays at least as $1/|\om|$ for high frequencies, and using assumption~\ref{as:causality} we find the usual Kramers-Kronig relations,
\begin{align}
\Re\,\chi(\om) &= \frac{1}{\pi}\cP\intwp \frac{\Im\,\chi(\om')}{\om'-\om}, \label{eq:KK1}\\
\Im\,\chi(\om) &= -\frac{1}{\pi}\cP\intwp \frac{\Re\,\chi(\om')}{\om'-\om},\label{eq:KK2}
\end{align}
where $\cP$ stands for the principal part. It follows from \ref{as:reality} that we can rewrite the first of these as
\begin{align}
\Re\,\chi(\om) = \frac{2}{\pi}\cP\intwhp \frac{\om'\Im\,\chi(\om')}{\om'^2-\om^2}.
\end{align}
Finally, if we consider the static response, we find that
\begin{align}\label{eq:diamagImpossible}
\chi(0) = \frac{2}{\pi}\cP\intwhp \frac{\Im\,\chi(\om')}{\om'},    
\end{align}
where we have omitted the real-part symbols, as an imaginary zero frequency response is prohibited by assumption~\ref{as:reality}. Here we arrive at the crux of the matter, as the right hand side of Eq.~\eqref{eq:diamagImpossible} contains only positive quantities, by the passivity assumption~\ref{as:passivity}. We must therefore conclude that $\chi(0) > 0$, based on the seemingly benign assumptions \ref{as:causality}-\ref{as:passivity}. This is clearly in conflict with diamagnetism where $\chi(0)<0$. 

We note that it is at times useful to work with permeability $\mu(\om) = \mu_\infty + \chi(\om)/[1-\chi(\om)]$, where $\mu_\infty$ is the medium response at infinite frequency. Diamagnetism is then the statement that
\begin{align}
    \mu(0)<1.
\end{align}
In this process, we introduce another physically justified assumption:
\begin{enumerate}
\setcounter{enumi}{3}
    \item \label{as:highfreq} Waves at wavelengths far smaller than the medium's typical atom-atom separation acts as if in vacuum. Using Maxwell's equations, we can show that this implies that $\mu \rightarrow 1$ as $\om\rightarrow \infty$. Physically, we can argue that this must be the case, as any medium must act as a series of small independent emitters rather than as a collective for small enough wavelengths. Therefore, the medium acts as free-space, i.e. transparent, to waves of a sufficiently high frequency.\footnote{An alternative way of considering this is that, at high enough frequencies, the wave is sufficiently off-resonance with any transition in consideration that the medium does not response. The conclusion of this line of reasoning is the same.}
\end{enumerate}
The requirement that $\mu(\om\rightarrow \infty) = 1$ is important also to preserve the commutation relations between the electric and magnetic fields when introducing material responses, and as such, is a fundamental requirement of the theory \cite{steveTRK}. We can arrive at this by using the sum rule where
\begin{align}\label{eq:opticalSumRule}
\Re\lel[\sum_{j} \mu(\om_j(\bk))v^j_p(\bk) v^j_g(\bk)\rer] = c^2,
\end{align}
for each $\bk$, where $\om_j(\bk)$ is the $j^\tm{th}$ polariton branch frequency defined through solving the dispersion relation $k^2= c^2\om^2n^2(\om)$ for $\om$ with $c$ being the speed of light. Here $v^j_p = \om_j/k$ and $v^j_g = \partial \om_j/\partial k$ are the associated phase and group velocities for each branch. Physically, there must exist a photon-like branch where $\om_j \sim   k$, for which the phase and group velocity approach $c$ as $k\rightarrow\infty$. This is simply because the photon-like branch must approach free-space. The other branches become increasingly matter-like in the same limit, with a group velocity that approaches zero. Hence, $\mu(\om)$ must approach unity at high frequencies. Using this, we can then rewrite Eq.~\eqref{eq:diamagImpossible} as
\begin{align}\label{eq:diamagImpossibleMu}
\mu(0) = 1 + \frac{2}{\pi}\cP\intwhp \frac{\Im\,\mu(\om')}{\om'} \geq 1,
\end{align}
which is in conflict with diamagnetism. As we shall outline below, the resolution to this conflict involves a combination of elements of previous work, together with new insights.

\section{Previous arguments}\label{sec:PrevArguments}
Multiple ways of resolving this enigma have been proposed over the years. In this section, we will here highlight the main schools of thought. While these schools of thought make interesting points, we will also argue that they do not resolve the issue. More details will given in Appendix~\ref{app:nonResolutions}. We can start by noting that, as diamagnetism is a ubiquitous phenomena, we believe that any explanation should not only resolve the contradiction between diamagnetism and the Kramers-Kronig relations, but also explain the physical mechanism by which the contradiction is resolved. Additionally, this explanation should not rely on the particulars of any particular medium, as diamagnetism is a widespread phenomenon. As mentioned in the introduction, the ideas proposed to resolve the conflict can be summarised as: 
\begin{itemize}
\item[(1)] that the magnetisation loses physical meaning at high frequencies,
\item[(2)] that a spatially non-local permeability $\mu(\om,\bk)$ is necessary,
\item[(3)] that diamagnetism is a nonlinear phenomenon,
\item[(4)] that the mathematics of the passivity assumption is too restrictive.
\end{itemize}
Let us first consider (1) and (2), as they are both arguments that aim to circumvent $\mu_\infty = 1$ in  Eq.~\eqref{eq:diamagImpossibleMu}, and thereby challenge the validity of the Kramers-Kronig relations. In the former \cite{landau}, an explicit high-frequency cut-off is introduced (or other similar modifications \cite{ModifiedKKSilveirinha2011}) in the Kramers-Kronig relations such that
\begin{align}\label{eq:diamagImpossibleMuHighFreq}
\mu(0) = \mu_a + \frac{2}{\pi}\cP\int_0^{\om_a} d\om'\; \frac{\Im\,\mu(\om')}{\om'} \geq 1,
\end{align}
where $\om_a$ is a high-frequency cut-off and $\mu_a = \mu(\om_a)$. Physical arguments set the $\om_a$ around the inverse length scale of the medium constituents (such as atoms), above which the permeability has no physical meadning. It is argued, thence, that $\mu_a$ does not necessarily carry physical meaning and can therefore be distinct from unity. Diamagnetism is then made possible by noting that only $\mu(0)-\mu_a \geq 0$ is implied by Eq.~\eqref{eq:diamagImpossibleMuHighFreq}. However, the same arguments can be made for any macroscopic media with a permittivity $\varepsilon$ and permeability $\mu$, as the macroscopic fields in general relies on that the wavelength is significantly larger than the medium constituents. In particular, it does not explain why the same argument does not apply to dielectric responses. Also, while it is true that the high-frequency permeability does not necessarily carry physical meaning by itself, the high-frequency cut-off $\om_c$ introduces a host of mathematical complications explored in, for instance, Ref.~\cite{norway_modifiedKK}. 

Assertion (2), that a spatially-dependent permeability $\mu(\om,\bk)$ is necessary, also aims to circumvent $\mu_\infty = 1$. In this case, the Kramers-Kronig relations can be formulated for each wavenumber $\bk$ such that the magnetostatic permeability must obey
\begin{align}\label{eq:nonlocalPermeability}
\mu(0,\bk) = \mu_\infty(\bk) + \frac{2}{\pi}\cP\int_0^\infty d\om'\; \frac{\Im\,\mu(\om',\bk)}{\om'}.
\end{align}
Indeed, given that $\Im\,\mu(\om,\bk) \geq 0$, then this implies that that $\mu(0,\bk) \geq \mu_\infty(\bk)$. Therefore $\mu(0,\bk)<1$ is possible if $\mu_\infty(\bk) < 1$. This approach is discussed in Refs.~\cite{pitaevskii2012, alusPaper, norway_diamagnetism_1} and others, where the focus is on spatially non-local conductors and metamaterials. It is plausible that such a model can be made consistent with the optical Thomas-Reiche-Kuhn sum rule in Eq.~\eqref{eq:opticalSumRule}, given that $\bk$-dependencies are appropriately introduced. However, care must be taken to ensure that the permeability $\mu(\om_j(\bk),\bk) \rightarrow 1$ as $k \rightarrow \infty$, as well as ensure appropriate behaviour for the phase $v_p^j$ and group velocities $v_g^j$. This can be a challenge, as spatially non-local effects are not necessarily restricted by causality, by their very nature.

More importantly, while it is possible that such a model can be constructed, the physical mechanism for the general and ubiquitous emergence of a spatially non-local permeability is unclear. Many regular, stationary, insulators are known to exhibit diamagnetic responses and spatially non-local effects typically involve movement and transport of the medium constituents, such as electron diffusion that requires a spatially non-local permittivity $\varepsilon(\om,\bk)$ \cite{AgranovichGinzburg}. We agree that diamagnetism through spatial non-locality is likely possible, but we argue that it is unlikely to be the predominant mechanism. Moreover, it suggests an explanation that is specific to the medium investigated as the spatial dependence of the permeability is determined by the structure of the medium.

In both (1) and (2), we want to note that the separation between the \textit{transverse} part of the polarisation $\Ph^\perp$ and magnetisation $\Mh^\perp$ is arbitrary and chosen by convention such that $\jh^\perp = \partial_t \Ph^\perp + \curl{}\Mh^\perp$. Only the \textit{longitudinal} component of the polarisation is directly defined through $\dive{}\Ph^{\parallel} = - \rho_b$ where $\rho_b$ is the bound charge density. In fact, as is discussed in, for example, Refs.~\cite{landau, AgranovichGinzburg} and is used more directly in Ref.~\cite{norway_diamagnetism_1}, we can exchange any transverse permittivities $\varepsilon_\perp$ and permeabilities $\mu$ with $\varepsilon_\perp'$ and $\mu'$ given that they fulfil
\begin{align}\label{eq:epsmuChange}
\varepsilon_\perp' + \frac{k^2 c^2}{\om^2}\left(1-\frac{1}{\mu'}\right) = \varepsilon_\perp + \frac{k^2 c^2}{\om^2}\left(1-\frac{1}{\mu}\right).
\end{align}
In that sense, neither the transverse permittivity nor the permeability has any unique physical meaning at any non-zero frequency. Any resolution of the diamagnetism-causality conflict, therefore, should not rely on this separation. This is because Maxwell's equations involve only the total current and not, individually, the polarisation and magnetisation degrees of freedom. This is something that is at the heart of a currently hotly debated gauge ambiguity in cavity QED \cite{adam1, nori1, myself}.\footnote{These occurs when energy-dependent approximations are made as any chosen separation between polarisation and magnetisation changes the energy distribution of the system.} As can be expected based on this, it is also possible to subsume all magnetic effects into a spatially non-local permittivity $\varepsilon(\om,\bk)$. This can be found in, for instance, \citet{landau} and \citet{AgranovichGinzburg}[p. 34]. While this can be useful for proving required properties, it does not provide a practical solution of how diamagnetism ought to be modelled, nor why it needs to be done in such a way.

The assertion in (3) that diamagnetism might be a nonlinear phenomenon is usually discussed in the context of macroscopic quantum electrodynamics \cite{scheel, scheelUnified}. 
In this case, we note that diamagnetism must be a phenomenon amenable to linear response theory, as experimental measurements show a linear dependence of the magnetisation with the applied magnetic field. Furthermore, we show that linear response theory is indeed applicable below.

We will discuss (4), that $\Im\;\chi(\om)\geq 0$ is too restrictive \cite{markel}, further in Section~\ref{sec:Resolution}. 
In short, we agree with this. Similarly to spatial non-locality, however, the fact alone that the assumption is too restrictive and that more general formulations are possible does not provide a concrete reason as to why this is required for diamagnetism, nor a physical mechanism as for how $\Im\;\chi(\om) < 0$ occurs. 

In conclusion, therefore, we believe that the previously proposed solutions focus on plausibility arguments, i.e. how diamagnetism \textit{could} be made consistent with causality and the Kramers-Kronig relations. They do not actually resolve the diamagnetism-causality conflict on a practical level as they do not present the precise mechanism by which causality is recovered. It is this crucial missing element that we provide.

\section{The resolution.}\label{sec:Resolution}
We will, in this section, provide a resolution to the diamagnetism-causality conflict. In particular, we will focus on the bulk media response of insulators and present a resolution that is fully compatible with macroscopic quantum electrodynamics without any modifications.

\subsection{Magnetostatics is different from the dynamics}
Let us start by noting that there are two \textit{physically distinct} $\om\rightarrow 0$ limits for magnetism in electrodynamics: the wave-like limit and the magnetostatic limit. This is well-known \cite{jackson, stratton}, but the limits are worth approaching separately here so we shall repeat the argument. We start with Maxwell's equations
\begin{align}\
\dive{}\Eh &= \rho, \label{eq:Maxwell} \\
\dive{}\Bh &= 0, \nonumber \\
\curl{}\Eh &= -\partial_t \Bh, \nonumber\\
\curl{}\Bh &= \partial_t \Eh + \jh, \nonumber
\end{align}
where we work throughout in units where $\varepsilon_0 = 1 = c = \hbar$. Let's suppose that $\jh = \delta\jh + \jh_\tm{ext} = \partial_t \Ph + \curl{}\Mh + \jh_\tm{ext}$ where 
\begin{align}
\Ph(t) &= \int_{-\infty}^t dt'\, \chi_E(t-t')\Eh(t'), \\
\Mh(t) &= \int_{-\infty}^t dt'\, \chi(t-t')\Bh(t')
\end{align}
are the medium polarisation and magnetisation, respectively, and $\jh_\tm{ext}$ is some external current. In frequency-space, we can rewrite the final Maxwell equation as
\begin{align}\label{eq:Maxwell_4rewrite}
\curl{}\lel[\Bh/\mu(\om)\rer] = -i\om \varepsilon(\om) \Eh + \jh_\tm{ext},
\end{align}
where we introduced the permittivity $\varepsilon(\om) = 1+\chi_E(\om)$ and the permeability $\mu(\om) = 1/(1-\chi(\om))$.
If we consider wave-dynamics, we can apply another $\curl{}$ and find
\begin{align*}
\curl{}\lel[\curl{}\Bh\rer]-\om^2\varepsilon(\om)\mu(\om)\Bh = \mu(\om)\curl{}\jh_\tm{ext}.
\end{align*}
Expanding $\Bh(\bx) = (2\pi)^{-3}\intk \exp(i\bk\cdd\bx)\Bh(\bk)$ results in the dispersion relation $k^2 = \om^2\varepsilon(\om)\mu(\om)$. Letting $\om\rightarrow 0$ now also demands that $k\rightarrow 0$, which we refer to as the wave-like limit. Clearly we must have that $\Im\,\varepsilon(\om)\mu(\om) > 0$ at all frequencies in order for the solution to describe absorption.

However, if we instead let $\om\rightarrow 0$ in Eq.~\eqref{eq:Maxwell_4rewrite}, then we find
\begin{align}
\curl{}\Bh = \mu(0)\jh_\tm{ext}.
\end{align}
This is the magnetostatic limit in which $\bk$ is not necessarily zero, akin to the electrostatic limit. It is also the limit where diamagnetism is relevant, and where we note that spatially non-local effects can play a role. Importantly for us, however, the magnetostatic limit poses no restrictions on the sign of $\Im\,\mu(\om)$, and by extension $\Im\,\chi(\om)$. 

Let us re-examine the relationship 
\begin{align}
\Mh(\om) &= \chi(\om)\Bh(\om)\leftrightarrow \Mh(t) = \int_{-\infty}^t dt'\,\chi(t-t')\Bh(t') 
\end{align}
in some detail. Based on this expression, we see that the sign of $\Im\,\chi(\om)$ determines whether the magnetisation $\Mh$ is lagging or leading in phase compared to the $\Bh$-field. The fact that $\Mh(t)$ decays in time is a consequence of causality in assumption~\ref{as:causality}, as the poles of $\chi(\om)$ are in the lower half-plane. As such, the sign of $\Im\,\chi(\om)$ does not determine the medium passivity, which it is commonly attributed to do (assumption~\ref{as:passivity}). We have arrived at the point made by \citet{markel}, albeit by a different route. As such, we argue that there \textit{must} be a negative part of $\Im\,\chi(\om)$ for diamagnetism to appear, and it only remains to show that this negative value does not, in fact, conflict with causality.

\subsection{Linear response theory yields negative $\Im\,\chi(\om)$}
It is not, however, enough that negative $\Im\,\chi(\om)$ is possible. We must explain how and why it appears. We should simultaneously ensure that $\Im\,\varepsilon(\om)\mu(\om) > 0$ is \textit{always} satisfied when $\mu(0) < 1$ for the solution to be physical and determine the physical phenomenon that ensures this behaviour. Furthermore, the optical Thomas-Reiche-Kuhn sum rules \cite{steveTRK}, along with physical considerations, require that both $\mu \rightarrow 1$ and $\varepsilon \rightarrow 1$ when $\om\rightarrow \infty)$. To resolve this issue, we need to accept that diamagnetism should not be considered in isolation from other electromagnetic phenomena.

\begin{figure}\centering
\includegraphics[width=0.85\textwidth]{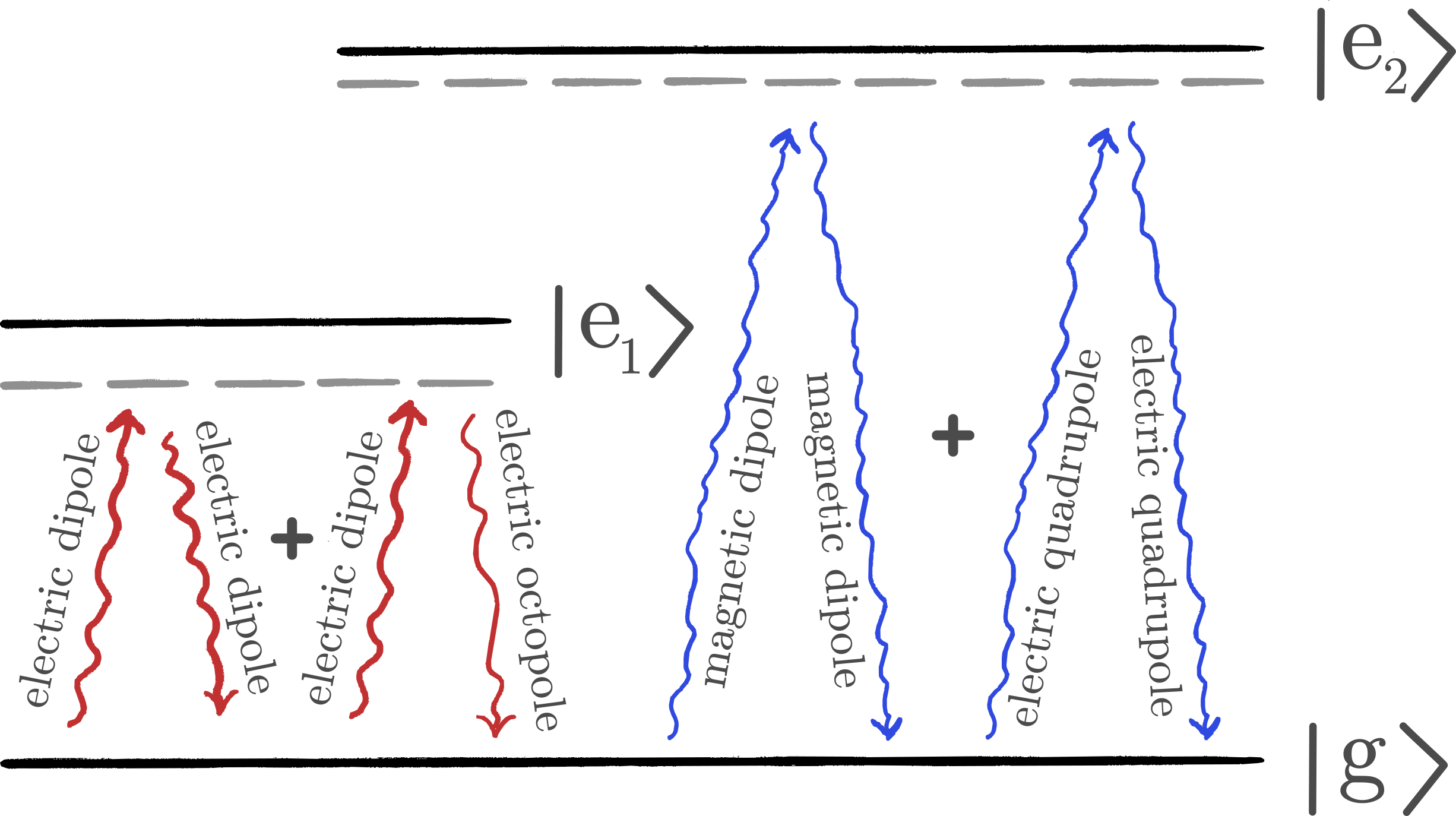}
\caption{Energy level diagram for two example, off-resonant, transitions. In this example, transitions between $\ket{g}$ and $\ket{e_1}$ can be mediated by \textit{both} the electric dipole operator, as well as the electric octopole operator. A dipole transition is, however, not possible between $\ket{g}$ and $\ket{e_2}$, which is dominated by a combination of the magnetic dipole and electric quadrupole operators. All these combinations are required for a self-consistent model of the diamagnetic response.}
\label{fig:transitions}
\end{figure}

We will centre our discussion around the different multipole orders of light-matter interaction and their characteristics \cite{jackson, stratton}: electric dipole, magnetic dipole, electric quadrupole, \textit{et cetera}. Each multiple order is related to certain angular momentum selection rules \cite{molecularQEDbook} where, for instance, the electric dipole interaction induce transitions between electronic states of the emitter that have a $\pm 1$ difference in angular momentum quantum number. For a linear susceptibility, at least two interactions between the emitter and the electromagnetic field are always required -- one to excite the emitter to a higher-laying energy level by absorbing a photon, and one to re-emit the light. The absorption and re-emission events need not necessarily involve the same multipole order, given that the selection rules are satisfied. This is illustrated in Fig.~\ref{fig:transitions}. We will, therefore, refer to a given contribution to the total current by their multipole order of interaction. For example, with the electric dipole-electric dipole susceptibility, we mean the induced current resulting from the emitter being excited through the electric dipole interaction, and the light being re-emitted by the same electric dipole interaction. In general, the magnetic susceptibility is a sum of spin and orbital angular momentum contributions. For simplicity, however, we will focus on the orbital contribution as the spin simply acts akin to a magnetic dipole.

With this in mind, the fact that diamagnetism appears at the same order as the electric quadrupole-quadrupole and the magnetic dipole-dipole response is well established (for instance \cite{molecularQEDbook, andrewsQuantumHamil, merlin}). This does not resolve the issue, however, as these contributions do not suffice to guarantee the positivity of $\Im\,\varepsilon(\om)\mu(\om)$ and, at the same time, respect the different selection rules of electric dipole, magnetic dipole and electric quadrupole transitions. The missing component is that \textit{the electric dipole-electric octopole and the electric dipole-magnetic quadrupole response} also enter at the same order and must be taken into account for consistency. 

This response naturally leads to a \textit{negative} $\Im\,\mu(\om)$ in some regions, while also ensuring that $\Im\,\varepsilon(\om)\mu(\om)$ is \textit{always positive}. The latter is guaranteed by the simple fact that the electric dipole-electric octopole/magnetic quadrupole response shares all resonances with the electric dipole-dipole response -- which is always significantly stronger -- leading to a reduction in the absorption rather than a gain for a wave propagating in the medium. There are, nevertheless, physical expectations and constraints associated with the electric dipole-electric octopole/magnetic quadrupole response. It should not generate a current at zero frequency and at infinite frequency it should generate a plasma-like response. We will here show how this behaviour is captured, and the argument is closed by the nature of the diamagnetic response itself.

Let us construct a minimal model for a current response that captures this behaviour. In particular, let us suppose that 
\begin{align}\label{eq:currentResponse}
\delta\jh(\om,\bk) = \bigg[&\sum_e \frac{\Delta_\tm{e-dip}^{eg} \om^2}{\om_{eg}^2-\om^2-2i\om \gamma_{e}} \nonumber\\
&-k^2\bigg(\frac{\Delta^{eg}_\tm{dia}}{\om_{eg}^2}-\frac{\Delta^{eg}_\tm{m-dip}}{\om_{eg}^2-\om^2-2i\om \gamma_{e}} \\
&\hspace{2.05cm} - \frac{\Delta^{eg}_\tm{quad}\om^2}{\om_{eg}^2-\om^2-2i\om \gamma_{e}}  + \frac{\Delta^{eg}_\tm{dip-oct}\om^2}{\om_{eg}^2-\om^2-2i\om \gamma_{e}} \bigg) \bigg]\Ah^\perp(\om,\bk),\nonumber
\end{align}
where $\Ah^\perp(\om,\bk)$ is the transverse part of the vector potential defined through $\Bh(\om,\bk) = i\bk\times\Ah^\perp(\om,\bk)$. In this, $g$ and $e$-labels denote ground and excited states $\ket{g}$ and $\ket{e}$, respectively, with $\om_{eg} = \om_e-\om_g$ being the transition frequency between the states with $\gamma_{e}$ being the linewidth of the excited state. Also, $\Delta^{eg}_i$ are the transition strengths associated with different transition types. These are proportional to the relevant transition moments, for instance with $\Delta^{eg}_\tm{e-dip} \propto \mb{d}^{ge}\cdd\mb{d}^{eg}$. Note that because the $\Delta^{eg}_i$ are proportional to the transition moments, only $\Delta^{eg}_\tm{e-dip}$ and $\Delta^{eg}_\tm{dip-oct}$ are both non-zero for the same excited state $\ket{e}$, as the other types of transitions have different selection rules \cite{molecularQEDbook}. We provide a derivation from first-principles in Appendix~\ref{app:linResponse}, and find that this form appears directly from linear response theory for the collection of emitters forming a medium. This derivation is by necessity lengthy and, while it provides a mathematical foundation, the physical insights can be arrived at independently of it, as we shall see below. 

It is natural to introduce a permittivity and permeability of the form
\begin{align}\label{eq:permittivityPermeability}
\varepsilon(\om) &= 1+\sum_e \frac{\Delta_\tm{e-dip}^{eg}}{\om_{eg}^2-\om^2-2i\om \gamma_{e}}, \\
\frac{1}{\mu(\om)} &= 1 + \sum_e \frac{\Delta^{eg}_\tm{dia}}{\om_{eg}^2}-\frac{\Delta^{eg}_\tm{m-dip}}{\om_{eg}^2-\om^2-2i\om \gamma_{e}}  - \frac{\Delta^{eg}_\tm{quad}\om^2}{\om_{eg}^2-\om^2-2i\om \gamma_{e}} + \frac{\Delta^{eg}_\tm{dip-oct}\om^2}{\om_{eg}^2-\om^2-2i\om \gamma_{e}}\nonumber.
\end{align}
\begin{figure}\centering
\includegraphics[width=0.7\textwidth]{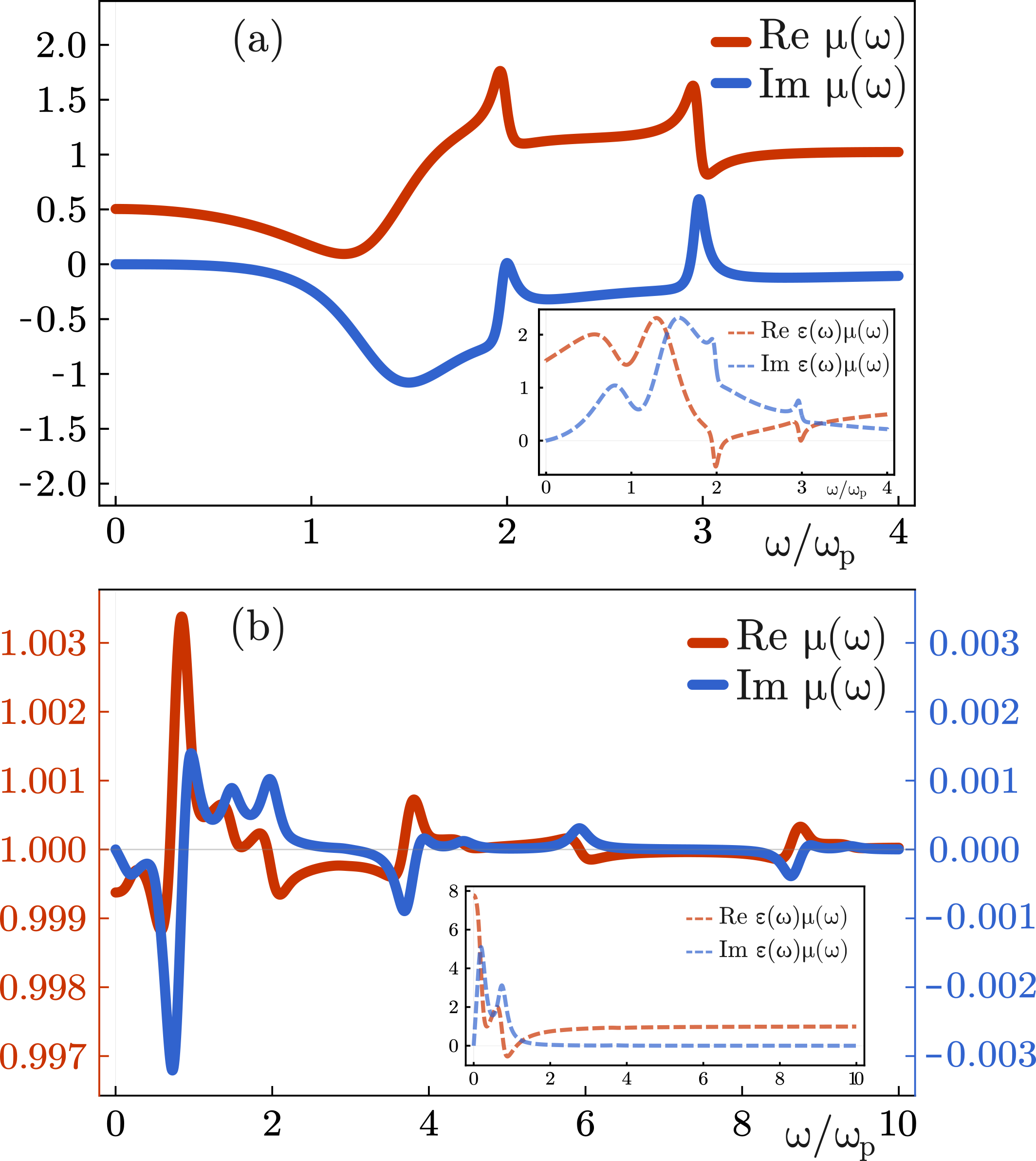}
\caption{(a) Illustrative example of the permeability $\mu(\om)$ for a strongly diamagnetic medium, using $\om_{eg}=\{1,2,3\}$, $\Delta^{eg}_\tm{e-dip} = \{2,0,0\}$, $\Delta^{eg}_\tm{m-dip} = \{0,1/16,0\}$, $\Delta^{eg}_\tm{quad} = \{0,0,1/64\}$, $\Delta^{eg}_\tm{dip-oct} = \{65/64,0,0\}$, $\Delta^{eg}_\tm{dia} = \{0,0,9\}$ and with $\gamma_{e} = \{0.18,0.05,0.04\}$, all in units of the plasma frequency $\om_p = \sqrt{\rho_0 q^2/m}$. Here $\rho_0$ is the emitter density of the medium, $q$ the electron charge and $m$ is the mass. The exact form of the transition strength definitions can be found in Appendix~\ref{app:rotAvg}, but we note that the coefficients are here chosen phenomenologically. Inset displays $\varepsilon(\om)\mu(\om)$ for the same medium. (b) Model atom-like medium as calculated from first principles, using all allowed transitions between $0<{n_x, n_y, n_z} \leq 6$, where $L_x = 0.2/2\pi \; \omega_p^{-1} = L_z$, $L_y = 0.4/2\pi \;\omega_p^{-1}$, $m = 500\,\omega_p$, $\rho_0 = 16000 \,\omega^3_p$ and $q = 1/\sqrt{32}$. Here all transitions have $\gamma_e = 0.15 \omega_p$. Note that this medium is weakly diamagnetic, as is expected for atoms.}
\label{fig:permeability}
\end{figure}
There are a few things we wish to note at this stage. First of all, in the $\om\rightarrow 0$ limit, we find that the permeability displays a competition between the paramagnetic and diamagnetic response, as all electric effects disappear and
\begin{align}
\mu(0) = \frac{1}{1+\sum_e (\Delta^{eg}_\tm{dia}-\Delta^{eg}_\tm{m-dip})/\om_{eg}^2}.
\end{align}
Hence a diamagnetic response appears when the paramagnetic response is suppressed, which is exactly as expected. This response is causal, as all the poles of 
\begin{align}
\chi(\om) &= 1-\frac{1}{\mu(\om)}\nonumber\\
&= \sum_e -\frac{\Delta^{eg}_\tm{dia}}{\om_{eg}^2} + \frac{\Delta^{eg}_\tm{m-dip}}{\om_{eg}^2-\om^2-2i\om \gamma_{e}} + \frac{\Delta^{eg}_\tm{quad}\om^2}{\om_{eg}^2-\om^2-2i\om \gamma_{e}} - \frac{\Delta^{eg}_\tm{dip-oct}\om^2}{\om_{eg}^2-\om^2-2i\om \gamma_{e}}
\end{align}
are in the lower half of the complex $\om$ plane, and thus $\Mh(t)$ decays in time. Furthermore, it is straightforward to show that $\Im\,\mu(\om)$ (and equivalently $\Im\,\chi(\om)$) becomes negative in frequency ranges determined by the electric dipole-octopole resonances, while $\Im\,\varepsilon(\om)\mu(\om)$ is strictly positive. The typical behaviour of the permeability, and its product with the permittivity, is illustrated in Figure~\ref{fig:permeability}(a) for some illustrative example values that fulfil all necessary conditions. This can be done, as our resolution does not require any particular knowledge about the material. We have, furthermore, computed the permeability and permittivity from first-principles in Figure~\ref{fig:permeability}(b), where we have chosen the particle-in-a-box as an example of atom-like, or model quantum dot, relative electron dynamics. As is well-known \cite{cohenQuantumMechanics}, the eigenfunction for the particle in a box are
\begin{align}
\psi(x,y,z) = \sqrt{\frac{8}{L_x L_y L_z}}\sin\left(\frac{n_x \pi}{L_x}\left[x+\frac{L_x}{2}\right]\right)\sin\left(\frac{n_y \pi}{L_y}\left[y+\frac{L_y}{2}\right]\right)\sin\left(\frac{n_z \pi}{L_z}\left[z+\frac{L_z}{2}\right]\right)
\end{align}
where $n_x$, $n_y$ and $n_z$ are the quantum numbers, $L_{x}$, $L_{y}$, $L_{z}$ are the dimensions of the box in the $x$, $y$ and $z$ directions, respectively. Here $\omega_{n_x n_y n_z} = (\pi^2/2m)(n_x^2/L_x^2+n_y^2/L_y^2+n_z^2/L_z^2)$ are the energies and we've assumed, for simplicity, that the states are equally long-lived. These can then be used to compute the multipole moments (explicitly defined in Eqns.~\eqref{eq:edip-edip}-\eqref{eq:mdip} in Appendix~\ref{app:linResponse}), and the rotational average can be done (Appendix~\ref{app:rotAvg}).

To show the positivity of $\Im\,\varepsilon(\om)\mu(\om)$ when $\Im\,\mu(\om)$ is negative, it is easiest rewrite $\Im\,\varepsilon\mu = |\mu|^2\Im\lel(\varepsilon/\mu^*\rer)$. We then consider $\om = \om_{eg}$ with $\ket{g}\leftrightarrow\ket{e}$ being an electric dipole-octopole allowed transition, as this is when $\Im\,\mu(\om)$ is maximally negative. Using Eq.~\eqref{eq:permittivityPermeability}, we then find that
\begin{align}
\Im\lel(\varepsilon/\mu^*\rer) \simeq \frac{1}{2\om_{eg}\gamma_{e}}\lel[\Delta^{eg}_\tm{e-dip}-\Delta^{eg}_\tm{dip-oct}\om_{eg}^2\rer] > 0 
\end{align}
where we have discarded the non-resonant terms. This is ensured to be positive by the nature of the multipole/long-wavelength expansion, as each subsequent term in the expansion must be rapidly decreasing in strength. Therefore $\Im\,\varepsilon\mu$ is, in fact, always positive.

Finally, we would like to note that this response can, of course, be written in terms of a spatially non-local permittivity $\varepsilon'(\om,\bk) = \varepsilon(\om) + (k^2/\om^2)\lel[1-1/\mu(\om)\rer]$ with $\mu' = 1$. While a non-local permittivity is perfectly valid, it introduces interpretation and computational complexity that is unnecessary. Importantly, however, we fulfil the requirement that $\Im\;\varepsilon'(\om,\bk) \geq 0$ for real $\om$ and $\bk$, as was proven, for instance, in Ref.~\cite{AgranovichGinzburg}.

\subsection{Sum rules enforce overall consistency}
Physical consistency demands compatibility both with physical expectations and the optical Thomas-Reiche-Kuhn sum rules \cite{molecularQEDbook}. Clearly, $\varepsilon(\om) \rightarrow 1$ as $\om\rightarrow \infty$. However, in this construction we require that
\begin{align}
\sum_e \frac{\Delta^{eg}_\tm{dia}}{\om_{eg}^2}+\Delta^{eg}_\tm{quad}-\Delta^{eg}_\tm{dip-oct} = 0
\end{align}
in order for $\mu(\om) \rightarrow 1$ as $\om\rightarrow \infty$. This is a sum rule and, as we prove in Appendix~\ref{app:sumRules}, is a direct consequence of the Thomas-Reiche-Kuhn sum rules \cite{molecularQEDbook} for the atomic transitions.\footnote{As a technical aside, this connection between the Thomas-Reiche-Kuhn sum rules and their optical counter-part is important for theoretical consistency. As we have shown here, it is possible to approach our expressions for the permittivity and permeability with knowledge of only one set of sum rules and derive the other.}

We should note the diamagnetic contribution to the total response is frequency-independent, which here appears as a requirement in order to satisfy the discussed physical constraints. This also has a physical explanation. The diamagnetic contribution appears as the difference between the kinetic and the canonical angular momentum carried by the atomic constituents. This contribution can be expressed as the angular momentum carried by the superposition of the charges' Coulomb field and the applied magnetic field $\Bh$. That is,
\begin{align}
\avg{\Lhh_\tm{dia}} = \intx \bx\times\lel[\Eh^\parallel\times\Bh\rer],
\end{align} 
when expressed in the Coulomb gauge, in which $\Eh^{\parallel}$ is purely the Coulomb field. This is proven in Appendix~\ref{app:momConv}, but a discussion can also been found in Ref.~\cite{cohen, steveSonjaAngularMomentum}. In other words, the diamagnetic contribution is essentially instantaneous, which implies frequency independence. In this sense, the diamagnetic contribution resembles a spatially non-local response, which is also instantaneous. The total response, is of course, still causal. This closes the argument.

\section{Conclusion.}\label{sec:Conclusion}
We have resolved the apparent conflict between diamagnetism and causality. Unlike earlier plausibility arguments, our analysis both proves the causality and identifies the physical mechanism that ensures it. The resolution relies on the observation that the electric dipole-electric octopole response enters at the same order as the diamagnetic, paramagnetic and electric quadrupolar responses, and must be included for consistency. This makes it possible for $\Im\,\chi(\om)$ to be negative in some frequency regions, while also guaranteeing that $\Im\,\varepsilon(\om)\mu(\om)$ is always positive, making the construction consistent with causality and the Kramers-Kronig relations. This appears directly in linear response theory given that all responses, at the same expansion order in the size of the emitter as compared to the wavelength, are taken into account (the long-wavelength expansion/multipolar expansion). 

We note that the condition, that the emitters are small compared to the wavelength, is also necessary for a collection of emitters to act as a medium, rather than individual emitters. The resolution is therefore guaranteed to work. Nonetheless, we assume a medium where the emitters are stationary (on the time-scale of the applied field) and expect that the formalism would need amended for moving emitters, even if the physics remains the same. The diamagnetic response of larger emitters is a different topic, which would be interesting to address in some later work.

Our focus here has been on formulating a physical explanation for diamagnetism and showing its consistency with causality, regardless of the underlying dynamics, applicable to all insulators. It would, now, be interesting to apply this to exotic and designed matter, such as metamaterials.  
We believe that our resolution provides a simple way forward based purely on the physics of diamagnetism. In terms of metamaterials, this only requires computing the multipole transition elements of the individual meta-atoms, which is already standard practice \cite{meinzer2014plasmonic}. Furthermore, we expect that similar arguments can be made to address the causality of diamagnetism in conductors when the applied field is very weak. However, since there is no binding energy, the magnetic response quickly becomes dominated by the formation of Landau levels \cite{landau1930diamagnetismus}, and hence the response becomes nonlinear \cite{NoltingRamakanth}. We leave the treatment of this to future work.

A consequence of the resolution of the diamagnetism-causality conflict is that it also explains the absence of dia-electric responses in regular, stationary, optical media. Any dia-electric response, where $\varepsilon(0) < 1$, would require that $\Im\,\varepsilon(\om) < 0$ at some frequency range while still guaranteeing the positivity of $\Im\,\varepsilon(\om)\mu(\om)$. That this is not possible in stationary media follows from the fact that the electric dipole enters alone in the long-wavelength expansion, without any mixed magnetic response that can play the equivalent role that the electric dipole-electric octopole response does for diamagnetism.

Finally, we note that this work extends macroscopic quantum electrodynamics as formulated in, for instance, Refs.~\cite{scheel, RobAndI} to include diamagnetic responses. This is given that the theory is formulated in such a way that it relies on $\Im\,\varepsilon(\om)\mu(\om)>0$ rather than requiring both $\Im\,\varepsilon(\om)$ and $\Im\,\mu(\om)>0$ independently.

\begin{acknowledgments}
The authors wish to acknowledge insightful discussions with James D. Cresser, as well as support from the Royal Commission for the Exhibition of 1851, the Royal Society (grant number RSRP/R/210005) as well as the UK Engineering and Physical Sciences Research Council under grant number EP/X033015/1.
\end{acknowledgments}

%\bibliography{diamagnetism}
%apsrev4-2.bst 2019-01-14 (MD) hand-edited version of apsrev4-1.bst
%Control: key (0)
%Control: author (72) initials jnrlst
%Control: editor formatted (1) identically to author
%Control: production of article title (-1) disabled
%Control: page (0) single
%Control: year (1) truncated
%Control: production of eprint (0) enabled
%

\clearpage
%\onecolumngrid

%\setcounter{equation}{0}
%\renewcommand{\theequation}{SI~\arabic{equation}}
%%\onecolumn
%\onecolumngrid

%\begin{center}
%    \Large Supplemental Information for \\ \textit{Is there a conflict between causality and diamagnetism?}
%\end{center}
\appendix

\section{Previously proposed resolutions}\label{app:nonResolutions}
\noindent We will here go into more details into the diamagnetism arguments as usually presented. In particular, we want to give more time to discuss the reasoning that leads to the arguments as presented. This list is not intended to be exhaustive, but rather to highlight the diverse approaches and ideas that have been proposed.

\subsection{The Landau-Lifshitz argument}\label{app:Landau}
\noindent Let us start with the school of thought put forward in \citet{landau}[p. 262], based around modifying the Kramers-Kronig relation: the magnetisation $\mb{M}$ becomes ill-defined above some frequency $\om_a$. This impacts both assumptions~\ref{as:causality} and \ref{as:highfreq}, and Landau and Lifshitz poses that this introduces a high-frequency cut-off to the integral in Eq.~\eqref{eq:diamagImpossibleMu}, as well as replaces $\mu_\infty = 1$ with $\mu(\om_a)\equiv\mu_a$, the latter of which can be less than unity. From Refs.~\cite{landau, ModifiedKKSilveirinha2011}, we then find
\begin{align}
  \label{eq:muLandau}
\mu(0) = \mu_a + \frac{2}{\pi}\cP\int_0^{\om_a} d\om' \; \frac{\Im\,\mu(\om')}{\om'},
\end{align}
which could indeed yield a diamagnetic response with $\mu(0)<1$. The argument that the magnetisation is ill-defined above some frequency is usually justified by considering a comparison of length-scales (here paraphrasing \citet{landau}): 
\begin{itemize}
\item[ ] We want to investigate the physics of the magnetisation, and thus consider a situation where any contribution from the polarisation $\bP$ is much smaller than a contribution from the magnetisation $\Mh$. For instance, let us work with the field inside a solenoid with a time-varying current applied, and suppose the body under consideration is placed along the solenoid's axis. We will here work with the usual definition of the magnetisation of a body, i.e. $\Mh = \intr \br \times \avg{\rho \mb{v}}$, where $\avg{\rho \mb{v}}$ is the mean current density. Furthermore, from consistency with the macroscopic Maxwell's equations, we have that
\begin{align}\label{eq:current}
\avg{\rho \mb{v}} = \partial_t\bP+\curl{}\Mh.
\end{align}
The electric field is this setting is only due to induction by the varying magnetic field, and we can estimate its strength using Maxwell's equation $\curl{}\Eh = -\partial_t\Bh$. This yields $E \sim \om l B$ where $l$ is the length of the body and $\om$ the frequency of variation. If we further put $\varepsilon_0\lel(\varepsilon - 1\rer) \sim \varepsilon_0$, then we get $\partial_t\bP\sim \om \varepsilon_0\Eh\sim \om^2 l \varepsilon_0 B = \om^2 l \varepsilon_0 B $. By definition, we also have $\mu_0\Mh = \chi \Bh$ and so $\curl{}\Mh \sim |\chi| B/\mu_0 l$. Therefore, for $\curl{}\Mh$ to be the dominant contribution to Eq.~\eqref{eq:current} we must have
\begin{align}
\om^2 \varepsilon_0 l \ll |\chi|/(\mu_0 l) \Rightarrow l^2 \ll |\chi| c^2/\om^2.
\end{align}
If we let $a$ be the atomic separation distance, then we must require that $l \gg a$ for the body to be macroscopic. Now, at some high frequency, we can have that $\om \sim c/a$, and so we find that
\begin{align}
l \ll \sqrt{|\chi|} a
\end{align}
which is clearly in contradiction with the assumption of a macroscopic body, given that $\chi$ does not grow arbitrarily large at high frequencies (which would also prohibit the formulation of the Kramers-Kronig relations). Ref.~\cite{landau} thus argues that the magnetisation must loose physical meaning at some high frequency $\om_a$.
\end{itemize}
This is true, however, at any frequency $\om$. As we discuss in the main text, the \textit{transverse} part of Eq.~\eqref{eq:current} is arbitrarily separated into polarisation and magnetisation contributions by convention, which is the part considered in the argument. The \textit{longitudinal} component of the polarisation is uniquely through $\dive{}\bP = - \rho_b$ where $\rho_b$ is the charge density. 
Also, the high-frequency cut-off does quite clearly not consider the contribution of the spin. Furthermore, this construction complicates the Kramers-Kronig relations, and much of the mathematical machinery needs to be modified. This was explored in Refs.~\cite{ModifiedKKSilveirinha2011, norway_modifiedKK} and leads to strong requirements on the asymptotic behaviour of the permeability.

\subsection{Spatial dispersion and metamaterials}\label{app:nonlocal}
\noindent There is another way to allow for $\mu_\infty \neq 1$, which is explored in Ref.~\cite{pitaevskii2012, norway_diamagnetism_1, norway_diamagnetism_2}. This is based on introducing spatial dispersion instead of making a high-frequency cut-off in the Kramers-Kronig relations. The argument goes as follows, paraphrased from Ref.~\cite{norway_diamagnetism_1}:
\begin{itemize}
\item[ ] Let us start with the macroscopic Maxwell's equations driven by an external current $\mb{J}_\tm{ext}$. If we make the ansatz of a spatial plane wave $\exp\lel( i \bk\cdd\br\rer)$, we find that Maxwell's equation yield
\begin{align}
\Bh &= \frac{i \mu_0\mu \bk \times \mb{J}_{\tm{ext},\perp}}{k^2-\om^2\varepsilon_\perp\mu/c^2} \label{eq:norwayB}\\
\Eh^\perp &= \frac{i \om \mu_0\mu \mb{J}_{\tm{ext},\perp}}{k^2-\om^2\varepsilon_\perp\mu/c^2} \label{eq:norwayE}\\
\Eh_\parallel &= \frac{\mb{J}_{\tm{ext},||}}{i\om\varepsilon_0\varepsilon_\parallel},
\end{align}
where we have to separate the longitudinal and transverse parts. We now require that the fields do not feel the impact of the medium, i.e. that the medium should act as free-space, in the $\om\rightarrow \infty$. This yields that $\varepsilon_\parallel \rightarrow 1$ and $\varepsilon_\perp \rightarrow 1$ when $\om\rightarrow\infty$. However, Ref.~\cite{norway_diamagnetism_1} note that Eqns.~\eqref{eq:norwayB} and \eqref{eq:norwayE} becomes independent of $\mu$ in the limit $\om \rightarrow \infty$ \textit{for a fixed $\bk$}, and therefore does not pose any restriction on $\mu$. This is different from what is referred to as `eigenmodal propagation' where $c^2 k^2 = \varepsilon_\perp \mu \om^2$, which would imply that $\varepsilon_\perp\mu \rightarrow 1$ when $\om\rightarrow\infty$. Non-eigenmodal propagation (i.e. a fixed $\bk$) requires the introduction of spatial dispersion in the medium such that the permeability becomes $\mu(\om,\bk)$. For instance, it is well known that the archetypal perfect diamagnet, that is, a superconductor, is highly non-local. This would allow for setting $\mu_\infty(\bk) \neq 1$ and diamagnetism is possible, whilst still being consistent with the regular Kramers-Kronig relations (see Eq.~\eqref{eq:nonlocalPermeability}. 
\end{itemize}
This argument raises a number of questions. As mentioned in the main text, it is not clear how the mechanism for spatial non-locality works physically. We would agree that that diamagnetism and spatial dispersion, in the electric quadrupole sense, enter at the same order of multipolar expansion in the microscopic Hamiltonian, and so it is feasible that they interact or have the same origin. Nonetheless, there are clear restrictions that need to be fulfilled, even with spatial dispersion in the medium, as discussed in the main text. 

\subsection{Macroscopic QED and the breakdown of linear response theory}\label{app:nonlinear}
\noindent The nonlinearity argument is based on the fact that the coupling term in the atomic Hamiltonian, that seems to be responsible for diamagnetism, is quadratic in both the atomic degree of freedom as well as in the magnetic induction. Thus linear response theory must break down. This argument can be found in many works on macroscopic quantum electrodynamics, and goes as follows \cite{scheelUnified}:
\begin{itemize}
\item[] As is well-known in cavity and molecular QED, the magnetic interaction terms in the atomic Hamiltonian \cite{molecularQEDbook, dispersion} is given by
\begin{align}\label{eq:magCoupling}
H^\tm{mag}_\tm{int} \simeq - \mh\cdd\Bh + \frac{e^2}{8 m}\lel(\mb{r}\times\Bh\rer)^2,
\end{align}
where $\mb{m}$, $\mb{r}$ and $m$ is the electron magnetic moment, coordinate and mass, respectively. We have here performed a multipolar expansion, valid for atomic length scales much smaller than the wavelength of light. The last term in Eq.~\eqref{eq:magCoupling} is commonly called the diamagnetic term, due to the sign, and is indeed bi-quadratic in the degrees of freedom. A medium formed from such atoms must therefore have a nonlinear reponse.
\end{itemize}
We would argue, however,that this does not preclude the use of linear response theory. First and foremost, the measured diamagnetic response is certainly linear in the applied magnetic field. Secondly, the diamagnetic response of a collection of atoms as commonly derived \cite{NoltingRamakanth} is consistent with linear response. The starting point is replacing $\br^2$ with $\avg{\br^2}$, as the magnetic induction is unlikely to cause an electronic transition in the interaction term $\lel(\mb{r}\times\Bh\rer)^2 \propto \mb{r}^2 \Bh^2$. Whilst the remaining term in the Hamiltonian is quadratic in $\Bh$ only, this leads to a \textit{linear} contribution to Maxwell's equations. It is noteworthy that it seems to be the fluctuations in $\mb{r}$ that drive the diamagnetic response, as $\avg{\br^2}$ is a measure of the fluctuations in electron position.

\subsection{Passivity}\label{app:passivity}
\noindent Another approach that has been suggested is that the passivity requirement in assumption~\ref{as:passivity} is too stringent. This has been proposed by \citet{markel} and another approach which nonetheless yields a qualitatively similar answer can be found in Ref.~\cite{norway_diamagnetism_1}. The idea is fairly straightforward, and summarised as \cite{markel, norway_diamagnetism_1}:
\begin{itemize}
\item[] If we allow $\Im \,\mu < 0$ for some frequencies $\om$ in Eq.~\eqref{eq:diamagImpossibleMu}, then the Kramers-Kronig would no longer be in conflict with diamagnetism. The argument put forward by \citet{markel} essentially boils down to a detailed investigation of the conditions under which $\Im \,\mu < 0$ does not yield exponentially growing waves. The quantity considered is the heating rate (note, in Gaussian units)
\begin{align}
q = \frac{\om}{8\pi}\lel[\Im\,\varepsilon(\om) \lel|\Eh(\om)\rer|^2+\Im\,\mu(\om) \lel|\Hh(\om)\rer|^2\rer],
\end{align}
and it is argued that for physically relevant scenarios (such as a magneto-dielectric sphere) we should always fulfil $\Im\,\varepsilon(\om) \lel|\Eh(\om)\rer|^2 \geq \Im\,\mu(\om) \lel|\Hh(\om)\rer|^2$. Furthermore, a second expression from the heating rate (again, in Gaussian units) is considered, where
\begin{align}
q_V = \frac{\om \lel|\Eh(\om)\rer|^2}{8\pi}\Im\lel[\varepsilon(\om)\mu(\om)\rer].
\end{align}
The $V$-subscript denotes the volume heating rate, and there is a surface heating rate that isn't important here. This then implies that we need not require that $\Im\,\varepsilon(\om) >0$ and $\Im\,\mu(\om)>0$ are individually satisfied but rather that $\Im\lel[\varepsilon(\om)\mu(\om)\rer] > 0$. This is indeed identical to the statement of only allowing exponentially decaying waves. 
\end{itemize}

We would also note that Ref.~\cite{norway_diamagnetism_1} argues a similar point based on considering Eq.~\eqref{eq:epsmuChange} in a pure dielectric where $\Im\,\varepsilon(\om) >0$. Using Eq.~\eqref{eq:epsmuChange}, Ref.~\cite{norway_diamagnetism_1} finds that $\Im\,\mu(\om)$ can be negative, given that
\begin{align}
\Im\,\mu(\om) > -\Im\,\varepsilon_\perp(\om)\frac{\om^2|\mu^2|}{k^2 c^2}.
\end{align}
We believe this is physically very similar to the argument put forward in the first part of Ref.~\cite{markel}. Or rather, they are a consequence of the same physical premise, i.e. considering the total current induced by the medium polarisation and magnetisations.

In this case, we fully concur with the observation that $\Im\lel[\varepsilon(\om)\mu(\om)\rer] > 0$ is a sensible requirement, as indeed waves must be exponentially decaying. It is simply that this does not by itself resolve the causality-diamagnetism conflict, and no physical mechanism is put forward to ensure that the above is fulfilled while the medium still exhibits a diamagnetic response. It is our work, which identifies the crucial role of the electric dipole-octopole contribution, that firmly establishes this point.

\clearpage
\section{Derivation from first principles}
The analysis presented in the main text showed that there is no conflict with causality if $\Im\,\mu(\om) <0$ for some range of frequencies and if it is also true that $\Im\,\varepsilon(\om)\mu(\om) > 0$ for all frequencies. The perceived mystery of the conflict between causality and the existence of diamagnetism is sufficiently long-standing that it is sensible, perhaps even necessary, to provide a fundamental proof of these conditions. To this end we provide, below,  a full derivation based on the well-established non-relativistic quantum electrodynamics of atoms interacting with fields, combined with the extension of this to the macroscopic quantum electrodynamics of electromagnetic fields interacting with macroscopic media. Our derivation is necessarily involved and technical, and we ask for the reader's patience. It is important, moreover, to emphasise the key points that we need to prove. These are:
\begin{itemize}
\item[(i)] Overall consistency requires the electric octopole contribution.
\item[(ii)] $\mu$ in the low frequency limit must display a competition between a diamagnetic and a paramagnetic contributions only, and the high-frequency limit must simultaneously yield both $\varepsilon(\om) \rightarrow 1$ and $\mu(\om) \rightarrow 1$ as $\om\rightarrow \infty$. Crucially, we need to answer why the sum rule \begin{align}
\sum_e \frac{\Delta^{eg}_\tm{dia}}{\om_{eg}^2}+\Delta^{eg}_\tm{quad}-\Delta^{eg}_\tm{dip-oct} = 0
\end{align}
appears and explain its physical significance.
\item[(iii)] Show that diamagnetism is linked with the difference between $\Lh_\tm{can}$ and $\Lh_\tm{kin}$.
\end{itemize}

We will start by examining a simple atom consisting of a pair of charges positions at $\br_\alpha$ with mass $m_\alpha$ and carrying charge $q_\alpha$. In particular, we will use that $q_1 = q$ and $q_2 = -q$ to form a charge-neutral system. We will then extend this to a continuum of oscillators forming a medium. 

\subsection{Diamagnetism in linear response theory}\label{app:linResponse}
As is typically done when modelling an optical medium, such as a dielectric, we will consider the current $\bj$ as induced by the an applied field in a continuous collection of atoms. For this, we will first calculate the induced current of the current operator for single atoms,
\begin{align}
\jhh(\bx,t) = \sum_\al q_\al \dot{\hat{\mathbf{r}}}_\al\delta(\bx-\rhh_\al),
\end{align}
in linear response theory. We then uniformly fill a region of space with a set of identical atoms at some density $\rho_0$ and sum up their total (and interfering) contributions to the current under the assumption that the atoms do not interact. 

We will here work in the Coulomb gauge, meaning that only $\Ah^\perp$ is dynamical. This choice does not impact any of the results or conclusions in the following, but allows us to easily treat all interactions on equal footing. To proceed, we first split the current operator into its canonical and the diamagnetic current contributions such that
\begin{align}
\jhh(\bx,t) &= \sum_\al \frac{q_\al}{m_\al}\lel[\phh_\al-q_\al \Ah^\perp(\rhh_\al,t)\rer]\delta(\bx-\rhh_\al) \nonumber\\
&= \jhh^\tm{can}(\bx)+\jhh^\tm{dia}(\bx).
\end{align}
Next we want to move to centre-of-mass and relative coordinates
\begin{align}
\bR &= \frac{m_1 \br_1 + m_2 \br_2}{m_1 + m_2}, \label{eq:centre-of-mass-Position}\\
\br &= \br_2-\br_1, \label{eq:relative-Position}
\end{align}
for which we also introduce the associated momenta
\begin{align}
\Phh &= \phh_1+\phh_2 \label{eq:centre-of-mass-Momentum}\\
\frac{\phh}{m} &= \frac{\phh_2}{m_2}-\frac{\phh_1}{m_1}, \label{eq:relative-Momentum}
\end{align}
where $\comm{\widehat{R}_i}{\widehat{P}_j} = i\delta_{ij}$ and $\comm{\hat{r}_i}{\hat{p}_j} = i\delta_{ij}$. Here $M = m_1 + m_2$ is the total mass, and we will also use the reduced mass $1/m = 1/m_1+1/m_2$. In order to keep the algebra to a minimum, we shall already at this stage make the assumption that $m_1 \gg m_2$ and let $m_1$ be large enough that dynamics is suppressed (all terms $\propto m_1/M \rightarrow 1$, $m_2/M \rightarrow 0$ and $1/M \rightarrow 0$). In this limit, each atoms' centre-of-mass is static (as would be expected in a stationary medium) and we can effectively treat $\Rhh$ as a classical variable. This yields well-localised set of atoms, which we can distribute evenly in space. These assumptions are strictly not necessary but are introduced here for the sake of simplicity (as centre-of-mass dynamics adds another layer of physics unrelated to the point we wish to make).

By inverting Eqns.~\eqref{eq:centre-of-mass-Position}-\eqref{eq:relative-Position} and Eqns.~\eqref{eq:centre-of-mass-Momentum}-\eqref{eq:relative-Momentum} and applying $m_1 \gg m_2$ and $m_1 \rightarrow \infty$ limits allows us to express the canonical and diamagnetic currents as
\begin{align}
\jhh^\tm{can}(\bx,\Rh) &= -\frac{q}{m}\;\ph\,\delta(\bx-\Rh-\rhh),\\
\jhh^\tm{dia}(\bx,\Rh) &= -\frac{q^2}{m}\;\Ah^\perp(\Rh+\rhh,t)\delta(\bx-\Rh-\rhh),
\end{align}
respectively. Here we have introduced $\Rh$ in the argument, alongside $\bx$ to signify that it now is a classical c-number coordinate. It also allows us to express the system Hamiltonian as
\begin{align}
\widehat{H} &= \sum_\al \frac{\lel[\phh_\al-q_\al \Ah^\perp(\rhh_\al,t)\rer]^2}{2m_\al} \rightarrow \frac{\lel[\phh+q\Ah^\perp(\Rh+\rhh,t)\rer]^2}{2m},
\end{align}
and the relevant interaction Hamiltonian is therefore 
\begin{align}
\widehat{H}_\tm{int}=\frac{q}{m}\phh\cdd\Ah^\perp(\Rh+\rhh,t).
\end{align}
We are now at a stage where we can apply linear response theory \cite{stevesBook}, where the leading order induced current by the applied field is given by $\avg{\delta\jhh(\bx)} = \avg{\delta\jhh^\tm{can}(\bx,\Rh)}+\avg{\delta\jhh^\tm{dia}(\bx,\Rh)}$ where
\begin{align}
\avg{\delta\jhh^\tm{can}} &= -i \int_{-\infty}^t dt'\; \bra{g}\comm{\jhh^\tm{can}(\bx,t)}{\hat{H}_\tm{int}(t')}\ket{g} \\
&= i \frac{q^2}{m^2}\int_{-\infty}^t dt'\; \bigg\langle g \bigg|\bigg[\phh(t)\delta\lel(\bx - \Rh - \rhh(t)\rer), \phh(t')\cdd\Ah^\perp(\Rh+\rhh(t'),t')\bigg]\bigg|g \bigg\rangle \nonumber\\
\avg{\delta\jhh^\tm{dia}} &= -\frac{q^2}{m}\bra{g}\Ah^\perp(\Rh+\rhh,t)\delta(\bx-\Rh-\rhh)\ket{g}
\end{align}
for some ground state $\ket{g}$ and where the operators are expressed in the interaction picture such that $\Ohh(t) = \exp\lel(i\hat{H}_0 t\rer)\Ohh\exp\lel(-i\hat{H}_0 t\rer)$ where $\hat{H}_0$ is the non-interacting part of the system Hamiltonian. 

As usual, we insert a resolution of identity $\mathbb{I} = \ket{g}\bra{g}+\sum_e \ket{e}\bra{e}$, where $\ket{e}$ are all the excited states of $\hat{H}_0$. We will therefore require both
\begin{align}
&\bra{g}\hat{p}_i(t)\delta(\bx-\Rh-\rhh(t))\ket{e} \text{ and } \\
&\bra{e}\hat{p}_j(t)A^\perp_j(\Rh+\rhh(t),t)\ket{g}
\end{align} 
as well as the same expressions where $\ket{e} \leftrightarrow \ket{g}$. In order to proceed, we shall expand in the long-wavelength approximation, consistently keeping up to quadratic order, i.e. $\mathcal{O}(|\rh|^2/\lambda^2)$. In other words, we will approximate
\begin{align}
\Ah^\perp(\Rh+\rhh,t) &\simeq \lel[1+\lel(\rhh\cdd\grad{R}\rer)+\frac{1}{2}\lel(\rhh\cdd\grad{R}\rer)^2\rer]\Ah^\perp(\Rh,t) \\
\delta(\bx-\Rh-\rhh) &\simeq \lel[1-\lel(\rhh\cdd\grad{R}\rer)+\frac{1}{2}\lel(\rhh\cdd\grad{R}\rer)^2\rer]\delta(\bx-\Rh)\nonumber
\end{align}
in the expressions for the current response $\avg{\delta\jhh}$. Let us also expand 
\begin{align}
\Ah^\perp(\Rh,t) = \frac{1}{2\pi}\intw \Ah^\perp(\Rh,\om)e^{-i\om t}
\end{align}
and use 
\begin{align}
e^{-i\om_{eg} t}\int_{-\infty}^t dt' \; e^{i\om_{eg} t}\Ah^\perp(\Rh,\om)e^{-i\om t'}=\frac{\Ah^\perp(\Rh,\om)e^{-i\om t}}{i\lel(\om_{eg}-\om-i\gamma_{e}\rer)}
\end{align}
where $\om_{eg} = \om_e-\om_g$ and we have introduced the small damping factor of $\gamma_{e}$ as is commonly done \cite{stevesBook}. A similar expression for when $g \leftrightarrow e$ follows.

All of this taken together yields the expressions
%\begin{widetext}
\begin{align}
\hat{j}_{i}^\tm{can}(\bx,\Rh) &= \frac{q^2}{m^2}\intwd e^{-i\om t}\sum_e \frac{1}{\om_{eg}-\om-i\gamma_{e}} \label{eq:canonicalCurrent} \\ 
&\hspace{4cm} \times \bigg\{\lel(p_i^{ge}-\lel(p_i r_k\rer)^{ge}\partial_k^R+\frac{1}{2}\lel(p_i r_k r_l\rer)^{ge}\partial^R_k\partial^R_l\rer)\delta(\bx-\Rh) \bigg\}  \nonumber \\
&\hspace{4cm} \times \bigg\{ \lel(p_j^{eg}+\lel(p_j r_m\rer)^{eg}\partial_m^R+\frac{1}{2}\lel(p_j r_m r_n\rer)^{eg}\partial^R_m\partial^R_n\rer)A^\perp_j(\Rh,\om) \bigg\} \nonumber\\
&\hspace{5cm} - \lel(e \leftrightarrow g\rer),  \nonumber\\
\hat{j}_i^\tm{dia}(\bx,\Rh) &= -\frac{q^2}{m}\intwd e^{-i\om t}\bigg\{\lel(1-r_k^{gg}\partial_k^R+\frac{1}{2}\lel(r_k r_l\rer)^{gg}\partial^R_k\partial^R_l\rer)\delta(\bx-\Rh) \bigg\} \nonumber\\
&\hspace{4cm}\times\bigg\{\lel(1+r_m^{gg}\partial_m^R+\frac{1}{2}\lel(r_m r_n\rer)^{gg}\partial^R_m\partial^R_n\rer)A^\perp_i(\Rh,\om)\bigg\}, \label{eq:diamagCurrent}
\end{align}
%\end{widetext}
where $O_i^{ge} = \bra{g}\widehat{O}_i\ket{e}$ denotes the transition element of the operator between ground and excited state. In using this expression, we will keep only terms that are up to $2^\tm{nd}$-order in the long-wavelength expansion. Furthermore, as we will aim to rotationally average, and not (for now) consider chiral molecules, we will omit the chiral terms. As a quick overview, we will label the terms by their nature as
%\begin{widetext}
\begin{align}
p_i^{ge} p_j^{eg}:& \tm{ (electric dipole)-(electric dipole) interaction}, \label{eq:edip-edip}\\
p_i^{ge} (p_j r_m)^{eg}:& \tm{ chiral -- omitted}, \label{eq:chiral}\\
p_i^{ge} (p_j r_m r_n)^{eg}:& \tm{ (electric dipole)-(electric octopole/magnetic quadrupole)}, \label{eq:eocto}\\
(p_i r_k)^{ge} (p_j r_m)^{eg}:& \tm{ (elec. quadrupole/mag. dipole)-(elec. quadrupole/mag. dipole)}, \label{eq:mdip}
\end{align}
%\end{widetext}
where we note that all the non-chiral terms are required for consistency. All other combinations (up to index relabelling) enter at a higher order and can be safely neglected. 

\subsubsection{Sum rules}\label{app:sumRules}
What remains is to match up the appropriate canonical and diamagnetic current terms, and identify the physics that each involves. For this, we will require the following:
%\begin{widetext}
\begin{align}
r_i^{eg} &= \frac{p_i^{eg}}{i m \om_{eg}} \label{eq:sumRule_1}\\
\delta_{ij} &= \frac{1}{m}\sum_e \frac{1}{\om_{eg}} p_i^{ge}p_j^{eg}-(e\leftrightarrow g), \label{eq:sumRule_2}\\
(r_k r_m)^{gg}\delta_{ij} &= \frac{1}{m}\sum_e \frac{1}{\om_{eg}}\big\{\lel[(p_i r_k)^{ge}+(p_k r_i)^{ge}\rer]\lel(p_j r_m\rer)^{eg} + \lel[(p_m r_i)^{ge}+(p_i r_m)^{ge}\rer]\lel(p_j r_k\rer)^{eg}\big\}  \nonumber\\
&\hspace{4cm} - \lel(e \leftrightarrow g\rer)  \label{eq:sumRule_3}\\
&= \frac{1}{m}\sum_e \frac{1}{\om_{eg}} p_i^{ge}\big\{\lel(p_j r_m r_n\rer)^{eg}+i \delta_{jm}r_n^{eg}\big\} -\lel(e \leftrightarrow g\rer) \nonumber\\
&= \frac{1}{m}\sum_e \frac{1}{\om_{eg}} \big\{\lel(p_i r_m r_n\rer)^{ge}+i \delta_{ik}r_m^{ge}\big\}p_j^{eg} -\lel(e \leftrightarrow g\rer),\nonumber
\end{align}
%\end{widetext}
which are all consequences of the Thomas-Reiche-Kuhn sum rule and repeated applications of $i \hat{p}_i/m~=~[\hat{r}_i,\widehat{H}_0]$ and the commutator $[\hat{r}_i,\hat{p}_j]=i\delta_{ij}$. After some tedious but straightforward algebra, this results in the current response (per atom) of:
%\begin{widetext}
\begin{align}
\avg{\delta\jhh(\bx,\Rh)} = &\avg{\delta\jhh^{(1)}(\bx,\Rh)}+\avg{\delta\jhh^{(2)}(\bx,\Rh)} +\avg{\delta\jhh^{(3)}(\bx,\Rh)}+\avg{\delta\jhh^{(4)}(\bx,\Rh)}
\end{align}
where 
\begin{align}\label{eq:dipoleReponse}
\avg{\delta\hat{j}_i^{(1)}(\bx,\Rh)} = \frac{q^2}{m^2}\intwd e^{-i\om t} & \delta(\bx-\Rh)A^\perp_j(\Rh,\om) \\
&\times \sum_e \lel\{\lel(\frac{1}{\om_{eg}-\om-i\gamma_{e}}-\frac{1}{\om_{eg}}\rer)p_i^{ge}p_j^{eg}-\lel(e\leftrightarrow g\rer)\rer\}\nonumber
\end{align}
is the electric dipole current response, and
\begin{align}\label{eq:quadrupoleResponse}
\avg{\delta\hat{j}_i^{(2)}(\bx,\Rh)} = \frac{q^2}{2m^2}\intwd & e^{-i\om t} \lel[\partial_k \delta(\bx-\Rh)\rer]\lel[\partial_m A^\perp_j(\Rh)\rer] \\
& \times \bigg[ \sum_e \bigg\{\lel(\frac{1}{\om_{eg}}-\frac{1}{\om_{eg}-\om-i\gamma_{e}}\rer) \nonumber\\
& \hspace{2.5cm}  \times \lel[\lel(p_i r_k\rer)^{ge}+\lel(p_k r_i\rer)^{ge}\rer]\lel(p_j r_m\rer)^{eg}-\lel(e\leftrightarrow g\rer)\bigg\} \nonumber\\
&\;\:\, + \sum_e \bigg\{\lel(-\frac{1}{\om_{eg}-\om-i\gamma_{e}}\rer) \nonumber\\
& \hspace{2.5cm}  \times \lel[\lel(p_i r_k\rer)^{ge}-\lel(p_k r_i\rer)^{ge}\rer]\lel(p_j r_m\rer)^{eg}-\lel(e\leftrightarrow g\rer)\bigg\} \nonumber\\
&\;\:\, + \sum_e \lel\{\frac{1}{\om_{eg}}\lel[\lel(p_i r_m\rer)^{ge}+\lel(p_m r_i\rer)^{ge}\rer]\lel(p_j r_k\rer)^{eg}-\lel(e\leftrightarrow g\rer)\rer\} \bigg]   \nonumber
\end{align}
contains the electric quadrupole-electric quadrupole and magnetic dipole-magnetic dipole current responses respectively, as well as the mixtures of these. The final term is the \textit{genuinely diamagnetic} contribution, which we note is frequency-independent. Finally, we have the electric dipole-electric octopole/magnetic quadrupole current responses
\begin{align}\label{eq:octopoleReponse}
\avg{\delta\hat{j}_i^{(3)}(\bx,\Rh)} &= -\frac{q^2}{2m^2}\intwd  e^{-i\om t} \delta(\bx-\Rh)\lel[\partial_k \partial_m A^\perp_j(\Rh,\om)\rer] \\
&\hspace{0.5cm} \times  \sum_e \lel\{\lel[ \lel(\frac{1}{\om_{eg}}-\frac{1}{\om_{eg}-\om-i\gamma_{e}}\rer)p_i^{ge}\lel(p_j r_k r_m\rer)^{eg}+i\delta_{jk}\frac{p_i^{ge}r_m^{eg} }{\om_{eg}}\rer]-\lel(e\leftrightarrow g\rer)\rer\},\nonumber \\
\avg{\delta\hat{j}_i^{(4)}(\bx,\Rh)} &= -\frac{q^2}{2m^2}\intwd  e^{-i\om t} \lel[\partial_k \partial_m \delta(\bx-\Rh)\rer]A^\perp_j(\Rh,\om) \\
&\hspace{0.5cm} \times  \sum_e \lel\{\lel[ \lel(\frac{1}{\om_{eg}}-\frac{1}{\om_{eg}-\om-i\gamma_{e}}\rer)\lel(p_i r_k r_m\rer)^{ge}p_j^{eg}+i\delta_{ik}\frac{r_m^{ge}p_j^{eg} }{\om_{eg}}\rer]-\lel(e\leftrightarrow g\rer)\rer\},\nonumber
\end{align}
%\end{widetext}
both of which contribute to the electric-dipole-electric octopole susceptibility.

It might be useful to confirm the nature of the different terms at this point, and we shall consider the electric quadrupole and magnetic dipole current responses as examples. First, we rewrite
\begin{align}
\lel[\lel(p_i r_k\rer)^{ge}+\lel(p_k r_i\rer)^{ge}\rer] = -i\om_{eg}m\lel(r_i r_k\rer)^{ge},
\end{align}
and define the quadrupole moment operator\footnote{For simplicity here, we allow it to have a trace.} as $\widehat{Q}_{ik} = -q \rhh_i\rhh_k$. Let us also rewrite
\begin{align}
\lel[\lel(p_i r_k\rer)^{ge}-\lel(p_k r_i\rer)^{ge}\rer] = \epsilon_{ikl}\lel(\rh \times\phh\rer)_l^{ge}.
\end{align}
and define the magnetic dipole moment operator as $\mhh = -q \rhh\times\phh/2m$. Starting with the first part of Eq.~\eqref{eq:quadrupoleResponse}, and consider the part that is symmetric under the exchange of $i$ and $k$ as well as $j$ and $m$, we find that it can be rewritten as
%\begin{widetext}
\begin{align}
\avg{\delta\hat{j}_{i,\tm{symm.}}^{(2)}(\bx,\Rh)} = \intwd & e^{-i\om t} \lel[\partial_k \delta(\bx-\Rh)\rer]\lel[\partial_m A^\perp_j(\Rh,\om)\rer] \\
& \times \sum_e \lel\{\lel(\frac{\om_{eg}}{4}-\frac{\om_{eg}^2}{4\lel(\om_{eg}-\om-i\gamma_{e}\rer)}\rer)Q_{ik}^{ge}Q_{jm}^{eg}-\lel(e\leftrightarrow g\rer)\rer\},\nonumber
\end{align}
which is indeed the electric quadrupole-electric quadrupole current response. Similarly, if we consider the second term of Eq.~\eqref{eq:quadrupoleResponse}, and the part that is anti-symmetric under the exchange of $i$ and $k$ as well as of $j$ and $m$, then we find
\begin{align}
\avg{\delta\jhh_\tm{anti-symm.}^{(2)}(\bx,\Rh)} = \intwd  e^{-i\om t} & \sum_e  \bigg\{\lel(\frac{1}{\om_{eg}-\om-i\gamma_{e}}\rer) \\
& \times \lel[\mh^{ge}\times\grad{R}\delta(\bx-\Rh)\rer]\lel[-\mh^{eg}\cdd\Bh(\Rh,\om)\rer]-\lel(e\leftrightarrow g\rer)\bigg\}\nonumber,
\end{align}
%\end{widetext}
which, of course, is the magnetic dipole-magnetic dipole current response written in more familiar terms. The other terms follow similarly.

\subsubsection{Atomic $\rightarrow$ medium response}\label{app:mediumReponse}
Finally, as we wish to model a medium, we must distribute space uniformly with atoms and sum over all contributions, which can by done by integrating over $\Rh$ and introducing the atom density $\rho_0$ where $\int d^3R \,\rho_0 = N$: 
\begin{align}
\jhh^\tm{tot}(\bx) = \int d^3R \;\rho_0\; \jhh(\bx,\Rh)
\end{align}
We should note that 
\begin{align}
\int d^3R \; \lel[\partial_k^R\partial_m^R\delta(\bx-\Rh)\rer]A^\perp_j(\Rh,\om) = \partial_k\partial_m A^\perp_j(\bx,\om) = \int d^3R \; \lel[\partial_k^R\delta(\bx-\Rh)\rer]\lel[\partial_m^R A^\perp_j(\Rh,\om)\rer],
\end{align}
where the second equality follows from $\partial_k^R\delta(\bx-\Rh)~=~-\partial_k^R\delta(\Rh-\bx)$ and $\int d^3R \; \grad{R}\delta(\Rh-\bx)f(\Rh) = -\grad{}f(\bx)$. The total current response $\avg{\delta\jhh^\tm{tot}(\bx)} = \avg{\delta\jhh_\tm{tot}^{(1)}(\bx)}+\avg{\delta\jhh_\tm{tot}^{(2)}(\bx)} +\avg{\delta\jhh_\tm{tot}^{(3)}(\bx)}+\avg{\delta\jhh_\tm{tot}^{(4)}(\bx)}$, to second order in the long-wavelength expansion, thus becomes
%\begin{widetext}
\begin{align}
\avg{\delta\hat{j}_{i}^\tm{tot}(\bx)} &= \intwd e^{-i\om t} \bigg[\alpha^\tm{e-dip}_{ij}(\om) + \bigg\{\alpha^\tm{quad}_{ikmj}(\om) + \alpha^\tm{m-dip}_{ikmj}(\om) \\
&\hspace{5.6cm} +\alpha^\tm{dia}_{ikmj} +\alpha^\tm{dip-oct}_{ikmj}(\om)\bigg\}\partial_k\partial_m\bigg]A^\perp_j(\bx,\om) \nonumber \\
&=  \intwd e^{-i\om t} \avg{\delta\hat{j}_{i}^\tm{tot}(\bx,\om)} \nonumber
\end{align}
where we have introduced the current susceptibility matrices
\begin{align}
\alpha^\tm{e-dip}_{ij}(\om) &= \frac{\rho_0 q^2}{m^2}\sum_e \lel(\frac{2}{\om_{eg}}\rer)\lel(\frac{\om^2}{\om_{eg}^2-(\om+i\gamma_{e})^2}\rer)p_i^{ge}p_j^{eg}, \label{eq:edipPol}\\
\alpha^\tm{quad}_{ikmj}(\om) &= \frac{\rho_0 q^2}{m^2}\sum_e \lel(-\frac{1}{\om_{eg}}\rer)\lel(\frac{\om^2}{\om_{eg}^2-(\om+i\gamma_{e})^2}\rer)\lel[(p_k r_i)^{ge}+(p_i r_k)^{ge}\rer](p_j r_m)^{eg},  \label{eq:equadPol}\\
\alpha^\tm{m-dip}_{ikmj}(\om) &= \frac{\rho_0 q^2}{m^2}\sum_e -\lel(\frac{\om_{eg}}{\om_{eg}^2-(\om+i\gamma_{e})^2}\rer)\lel[(p_k r_i)^{ge}-(p_i r_k)\rer](p_j r_m)^{eg},  \label{eq:mdipPol}\\
\alpha^\tm{dia}_{ikmj}(\om) &= \frac{\rho_0 q^2}{m^2}\sum_e \lel(\frac{1}{\om_{eg}}\rer)\lel[(p_i r_m)^{ge}+(p_m r_i)^{ge}\rer](p_j r_k)^{eg},  \label{eq:diaPol} \\
\alpha^\tm{dip-oct}_{ikmj}(\om) &= \frac{\rho_0 q^2}{m^2}\sum_e \lel(\frac{1}{\om_{eg}}\rer)\lel(\frac{\om^2}{\om_{eg}^2-(\om+i\gamma_{e})^2}\rer)\lel[p_i^{ge}(p_j r_k r_m)^{eg}+p_j^{ge}(p_i r_k r_m)^{eg}\rer],  \label{eq:edipOctPol} 
\end{align}
%\end{widetext}
which account for the the electric dipole susceptibility, electric quadrupole susceptibility, magnetic dipole susceptibility, diamagnetic susceptibility as well as the electric dipole-octopole susceptibility, respectively. To simplify the expressions, we used the fact that we can always choose wavefunctions such that $\mb{r}^{ge} = \mb{r}^{eg}$ and $\mb{p}^{ge} = -\mb{p}^{eg}$. 

These are still at a level of current susceptibilities, and we've not yet split them between a permittivity $\varepsilon$ and a permeability $\mu$. For reference, they enter in Maxwell's equations as
\begin{align}\label{eq:maxwellSmall}
\curl{}\Bh(\bx,\om)= -i\om\Eh(\bx,\om) +\avg{\delta\jhh^\tm{tot}(\bx,\om)}.
\end{align} 
As discussed, there is a choice involved in this and it proves simpler to consider the limiting behaviour first.

The key point is that only a competition between the diamagnetic and paramagnetic response remains in the $\om\rightarrow 0$ limit. In particular,
\begin{align}
\avg{\delta\hat{j}^\tm{tot}_i(\bx,\om\rightarrow 0)} = \bigg[\alpha_{ikmj}^\tm{m-dip}(0) &+ \alpha_{ikmj}^\tm{dia}\bigg]\partial_k\partial_m A^\perp_j(\bx,0),
\end{align}
where $\alpha_{ikmj}^\tm{m-dip}(0)$ and $\alpha_{ikmj}^\tm{dia}$ have opposite signs, and where we remind the reader that $\alpha_{ikmj}^\tm{dia}$ does not have a frequency dependence. This must be the case, as a time-independent vector potential should not generate electric effects but the spatial structure can generate magnetic currents. Furthermore, we must also consider the $\om\rightarrow\infty$ limit, as the medium must then return to a free-space response. This can be done by taking the $\om\rightarrow\infty$ limit in the above expressions for the current susceptibility and then applying the sum rules in Eqns.~\eqref{eq:sumRule_2}-\eqref{eq:sumRule_3} in reverse. Let us, for this, define
\begin{align}
\alpha_{ikmj}^\tm{tot}(\om) = \alpha_{ikmj}^\tm{quad}(\om)+\alpha_{ikmj}^\tm{m-dip}(\om)+\alpha_{ikmj}^\tm{dia}+\alpha_{ikmj}^\tm{dip-oct}(\om).
\end{align} 
We then find
\begin{align}
\alpha_{ijkl}^\tm{tot}(\om\rightarrow\infty)=\sum_e \frac{1}{\om_{eg}}\bigg\{&\lel[(p_i r_k)^{ge}+(p_k r_i)^{ge}\rer](p_j r_m)^{eg} + \lel[(p_i r_m)^{ge}+(p_m r_i)^{ge}\rer](p_j r_k)^{eg} \nonumber\\
&- \lel[p_i^{ge}\lel(p_j r_k r_m\rer)^{eg}+p_j^{ge}\lel(p_i r_k r_m\rer)^{eg}\rer]\bigg\} = 0.
\end{align} 
It is, however, more straightforward to evaluate by taking a step back noting that all \textit{canonical} current contributions [Eq.~\eqref{eq:canonicalCurrent}] vanish in this limit, so $\jhh^\tm{can}(\bx,\Rh,\om) \rightarrow 0$ as $\om\rightarrow \infty$ as they scale at least as $1/\om$. What remains is the \textit{diamagnetic} current in Eq.~\eqref{eq:diamagCurrent}. However, after integrating over all centre-of-mass $\Rh$, only the plasma-like response from the dipole-dipole interaction
\begin{align}
\avg{\jhh^\tm{tot}(\bx,\om\rightarrow\infty)} = -\frac{q^2 \rho_0}{m}\Ah^\perp(\bx,\om\rightarrow\infty)
\end{align}
remains and all other contributions destructively interfere. This is typical for any dielectric, and still results $\varepsilon \rightarrow 1$ as the permittivity comes with an additional factor of $1/\om^2$. This proves points (i) and (ii) as set out at the start, as the electric-dipole-electric octopole response is required for the sum rules. We therefore find that the sum rule proposed in the main text [Eq.~(21)] is a consequence of the underlying Thomas-Reiche-Kuhn sum rules.

\subsubsection{Rotationally averaged response}\label{app:rotAvg}
We can now recover to the expression found in the main document proposed earlier by rotational averaging \cite{molecularQEDbook} and defining the $\Delta^{eg}_\tm{e-dip}$, $\Delta^{eg}_\tm{m-dip}$, $\Delta^{eg}_\tm{dia}$, $\Delta^{eg}_\tm{quad}$, $\Delta^{eg}_\tm{dip-oct}$ from the averaged transition moments. This is done by letting 
\begin{align}
\overline{p_i^{ge}p_j^{eg}} &= \delta_{ij}\ph^{ge}\cdd\ph^{eg}/3, \\
\overline{\alpha^\tm{tot}_{ikmj}} &= \mathcal{I}^{(4)}_{ikmj\mu\nu\gamma\delta}\alpha^\tm{tot}_{\mu\nu\gamma\delta}
\end{align} 
and
\begin{align}
\mathcal{I}^{(4)}_{ikmj\mu\nu\gamma\delta}=\frac{1}{30}\bigg(\begin{array}{ccc}
\delta_{ik}\delta_{mj}, & \delta_{im}\delta_{kj}, & \delta_{ij}\delta_{km}
\end{array} \bigg) \lel(\begin{array}{rrr}
4 & -1 & -1 \\
-1 & 4 & -1 \\
-1 & -1 & 4 
\end{array} \rer) \lel(\begin{array}{c}
\delta_{\mu\nu}\delta_{\gamma\delta} \\
\delta_{\mu\gamma}\delta_{\nu\delta} \\
\delta_{\mu\delta}\delta_{\nu\gamma}
\end{array} \rer).
\end{align}
We can now define 
\begin{align}
\chi_E(\om) &= \frac{\alpha^\tm{e-dip}_{i i}(\om)}{3\om^2}, \\
\chi(\om) &= -\frac{1}{30}\lel[4\alpha^\tm{tot}_{\delta\nu\nu\delta}(\om)-\alpha^\tm{tot}_{\nu\delta\nu\delta}(\om)-\alpha^\tm{tot}_{\nu\nu\delta\delta}(\om)\rer], \\
\zeta(\om) &= \frac{1}{30}\lel[4\alpha^\tm{tot}_{\nu\delta\nu\delta}(\om)-\alpha^\tm{tot}_{\nu\nu\delta\delta}(\om)-\alpha^\tm{tot}_{\nu\delta\delta\nu}(\om)\rer],\\
\upsilon(\om) &= \frac{1}{30}\lel[4\alpha^\tm{tot}_{\nu\nu\delta\delta}(\om)-\alpha^\tm{tot}_{\nu\delta\nu\delta}(\om)-\alpha^\tm{tot}_{\delta\nu\nu\delta}(\om)\rer],
\end{align}
through which we can further define the transition strengths as
\begin{align}
\Delta^{eg}_\tm{e-dip}/\om_p^2 &= \big(\mb{p}^{ge}\cdd\mb{p}^{eg}/m\om_{eg}\big), \\
\Delta^{eg}_\tm{quad}/\om_p^2 &= - \lel(\frac{1}{m\om_{eg}}\rer)\avgg{\big[(p_k r_i)^{ge}+(p_i r_k)^{ge}\big](p_j r_m)^{eg}}, \\
\Delta^{eg}_\tm{m-dip}/\om_p^2 &= - \lel(\frac{\om_{eg}}{m}\rer)\avgg{\big[(p_k r_i)^{ge}-(p_i r_k)^{ge}\big](p_j r_m)^{eg}}, \\
\Delta_\tm{dia}^{eg}/\om_p^2 &=  \lel(\frac{\om_{eg}}{m}\rer)\avgg{\big[(p_i r_m)^{ge}+(p_m r_i)\big](p_j r_k)^{eg}}, \\
\Delta_\tm{dip-oct}^{eg}/\om_p^2 &= \lel(\frac{1}{m\om_{eg}}\rer)\avgg{\big[p_i^{ge}(p_j r_k r_m)^{eg}+p_j^{ge}(p_i r_k r_m)^{eg}\big]}, 
\end{align}  
where we have used the polarisabilities defined in Eqns~\eqref{eq:edipPol}-\eqref{eq:edipOctPol}. Here $\om_p = \sqrt{\rho_0 q^2/m}$ is the plasma frequency for a medium of density $\rho_0$, and
\begin{align}
\avgg{f_{ikmj}} = -\frac{1}{30}\lel[4f_{\delta\nu\nu\delta}-f_{\nu\delta\nu\delta}-f_{\nu\nu\delta\delta}\rer],
\end{align} 
denotes the rotational averaging. This allows us to rewrite Eq.~\eqref{eq:maxwellSmall} as 
\begin{align}
\bigg\{-\lel[1-\chi(\om)\rer]\nabla^2-\om^2\lel[1+\chi_E(\om)\rer]\bigg\}\Ah^\perp(\bx,\om)=\lel[\zeta(\om)+\upsilon(\om)-1\rer]\grad{}\lel[\dive{}\Ah^\perp(\bx,\om)\rer].
\end{align}
Since we are working in Coulomb gauge, and $\dive{}\Ah^\perp = 0$, this can be written in the form of the current response
\begin{align}
\delta\jh(\om,\bk) = \bigg[&\sum_e \frac{\Delta_\tm{e-dip}^{eg} \om^2}{\om_{eg}^2-\om^2-2i\om \gamma_{e}} \nonumber\\
&-k^2\bigg(\frac{\Delta^{eg}_\tm{dia}}{\om_{eg}^2}-\frac{\Delta^{eg}_\tm{m-dip}}{\om_{eg}^2-\om^2-2i\om \gamma_{e}} \\
&\hspace{2.05cm} - \frac{\Delta^{eg}_\tm{quad}\om^2}{\om_{eg}^2-\om^2-2i\om \gamma_{e}}  + \frac{\Delta^{eg}_\tm{dip-oct}\om^2}{\om_{eg}^2-\om^2-2i\om \gamma_{e}} \bigg) \bigg]\Ah^\perp(\om,\bk),\nonumber
\end{align}
as discussed in the main text.

\subsection{Diamagnetism as angular momentum conservation}\label{app:momConv}
Let us consider the terms that appear in the dynamics of the angular momentum. Our starting point will here be the minimal-coupling action 
\begin{align}
    S = \int dt\; \sum_\alpha &\frac{m_\alpha}{2}\dbral^2+q_\al \dbral\cdd\Ah(\bral,t) - q_\al \phi(\bral,t) \nonumber\\
    &+q_\al\pd{\chi}{t}(\bral,t)+q_\al\dive{\bral}\lel[\dbral\chi(\bral,t)\rer],
\end{align}
where we have also added a gauge-fixing term $\chi(\bral,t)$ which we shall specify only at a later stage. Note that we have chosen not to work with a multipolar gauge, as we wish to work with a gauge that is globally valid, particularly when we extend the system to include multiple charge pairs \cite{cohen}[p. 331-332].\footnote{The Power-Zienau-Woolley transform is perfectly valid globally but is not necessarily a gauge choice \cite{cohen, Woolley}.} As we will not be making any energy-level truncations here, the gauge choice makes little impact but we want to emphasise that the following is valid not only in the Coulomb gauge but also the Poincar\'e gauge and the Lorenz gauge. In other words, we will not work with any gauge that explicitly references the charge system. By construction, we can now define the updated fields $\Ah'(\bral,t)=\Ah(\bral,t)+\grad{}\chi(\bral,t)$ and $\phi'(\bral,t)=\phi(\bral,t)-\partial_t\chi(\bral,t)$ and work with the primed coordinates instead, akin to Ref.~\cite{babiker1983derivation, RobAndI}. For notational convenience, however, we shall drop the prime moving forward. As expected, a variation of the action leads to the Lorentz force law
\begin{align}
    \ddbral = q_\al\lel[\Eh(\bral,t)+\dbral\times\Bh(\bral,t)\rer],
\end{align}
where the electric and magnetic fields are given as $\Eh= -\partial_t\Ah-\grad{}\phi$ and $\Bh=\curl{}\Ah$, respectively.

In Noether's theorem, we will now consider the infinitesimal rotation $R\bral\simeq\bral+\bs{\theta}\times\bral$, under which the coordinates and fields transform as
\begin{align}
    \delta\bral &= \thh\times\bral, \quad \delta\dbral=\thh\times\dbral,\nonumber\\
    \delta\Ah(\bral,t)&=\thh\times\Ah(\bral,t)+\lel[\lel(\thh\times\bral\rer)\cdd\grad{}\rer]\Ah(\bral,t), \nonumber\\
    \delta\phi(\bral,t)&=\lel(\thh\times\bral\rer)\cdd\grad{}\phi(\bral,t)
\end{align}
We can now proceed to vary the action in accordance with Noether's theorem \cite{noether}. Using the equations of motion, as well as the fact that the variation of the action is non-zero due to the presence of non-conservative forces, we find the dynamical equations for the angular momentum
\begin{align}\label{eq:angularMomentum}
    \sum_\al &\dd{}{t}\lel[m_\al \bral\times\dbral+q_\al\bral\times\Ah\rer] =\\
    &\sum_\al q_\al\dbral\times\Ah+\bral\times\grad{}\lel[\dbral\cdd\Ah-\phi\rer].\nonumber
\end{align}
Here we can identify three terms:
\begin{align}
    \bL_\tm{kin} &= \sum_\al m_\al \bral\times\dbral,\\
    \bL_\tm{dia} &= \sum_\al q_\al \bral\times\Ah(\bral,t), \\
    \bs{\tau} &= \sum_\al q_\al\bigg\{\dbral\times\Ah(\bral,t) +\bral\times\grad{}\lel[\dbral\cdd\Ah(\bral,t)-\phi(\bral,t)\rer]\bigg\}, 
\end{align}
representing the \textit{kinetic} angular momentum $\bL_\tm{kin}$, the diamagnetic angular momentum $\bL_\tm{dia}$ and the torque $\bs{\tau}$, respectively. Let us start by noting that this can be rewritten as
\begin{align}
    \dd{\bL_\tm{kin}}{t} &= \bs{\tau} - \dd{\bL_\tm{dia}}{t} \\
    &= \sum_\al q_\al \bral\times\lel[\Eh(\bral,t)+\dbral\times\Bh(\bral,t)\rer],\nonumber
\end{align}
which is simply the torque generated by the Lorentz force law, as must be the case. Furthermore, the \textit{canonical} angular momentum is simply
\begin{align}
    \bL_\tm{can}=\bL_\tm{kin}+\bL_\tm{dia} = \sum_\al \bral\times\bp_\al,
\end{align}
as $\bp_\al = m_\al \dbral + q_\al\Ah(\bral,t)$, whose dynamics is governed by
\begin{align}
\dd{\bL_\tm{can}}{t} = \sum_\al & \frac{q_\al}{m_\al}\bigg\{\bp_\al\times\Ah + \bral\times\grad{\bral}\lel[\bp_\al \cdd \Ah - \phi - q_\al \Ah^2\rer] \bigg\},
\end{align}
where we used that $\Ah\times\Ah = \mb{0}$. Diamagnetism, which enters in the final $\Ah^2$-term when the dynamics is written in canonical variables, appears to be nonlinear (as discussed in Refs.~\cite{scheel, scheelDiamag} as an argument for precluding linear response theory). This does not preclude the use of linear response theory, however.

What is significant here is that the diamagnetic contribution to the angular momentum is exactly the difference between the kinetic and canonical angular momentum, which has at times been referred to as a `hidden' momentum. This is significant for the similar reasons as to why it is significant for the resolution of the Abraham-Minkowski dilemma for the linear momentum associated with light in media \cite{steveAbrahamMinkowski}. In order for the angular momentum to be conserved, there must therefore exist an equal and opposite contribution to the angular momentum `hidden' in the field angular momentum. This has previously been discussed in Ref.~\cite{steveSonjaAngularMomentum} in the context of a static magnetic field. It echoes the discussion put forward by \citet{TiggelenLkin, Tiggelen2} in a different context, which uses the fact that diamagnetism can be seen as a manifestation of Lenz's law, stating that the current induced by a magnetic field produces an opposing field \cite{lenz}. That this contribution appears instantaneous is a consequence of the instantaneous nature of the Coulomb field. We can extend this to time-varying fields by rewriting the diamagnetic angular momentum as
\begin{align}
\bL_\tm{dia} &= \sum_\al q_\al \bral\times\Ah(\bral,t) \nonumber\\
&= \intx \rho(\bx)\bx\times\Ah(\bx,t),
\end{align}
where $\rho(\br) = \sum_\al q_\al \delta(\br-\bral)$ is the charge density. Using Gauss' law $\dive{}\Eh = \rho$ and integration by parts, we can rewrite the diamagnetic angular momentum as
\begin{align}
\bL_\tm{dia} = \intx \bx\times\lel[\Eh^\parallel\times\Bh\rer] - \intx \lel\{\Eh^\parallel\times\Ah+E^\parallel_i\lel(\bx\times\grad{}\rer)A_i\rer\}.
\end{align}
The first term is the angular momentum associated with the total longitudinal electric field $\Eh^\parallel$ overlapping with the \textit{applied} (transverse) magnetic field $\Bh$. The second term subtracts the spin and orbital angular momentum carried by the overlap between the total longitudinal field ($\Eh^\parallel$) and the longitudinal components of the applied field ($\Ah^\parallel$). Therefore, the diamagnetic angular momentum contribution originates from the Coulomb field component of $\Eh^\parallel$ overlapping the applied field. In the Coulomb gauge, where $\Ah$ is completely transverse, this last term is identically zero but it persists in other gauges (such as Lorenz gauge).  Note the lack of any time-delayed response, meaning that the effect appears instantaneous. As this involves a split-up between longitudinal and transverse fields, this shouldn't be a surprise as only their sum is typically causal \cite{stratton, jackson}.

Finally, as a check, let us confirm that this diamagnetic angular momentum contribution indeed reproduces the known expressions for the diamagnetic magnetisation. For this, we need to switch to centre-of-mass and relative coordinates, defined as before as
\begin{align}
\bR &= \frac{m_1 \br_1 + m_2 \br_2}{m_1 + m_2}, \\
\br &= \br_2-\br_1,
\end{align}
which can be inverted as $\br_1 = \bR-(m_2/M)\br$ and $\br_2 = \bR+(m_1/M)\br$ where $M = m_1 + m_2$ is the total mass. We will also use the reduced mass $m^{-1} = m_1^{-1}+m_2^{-1}$. To proceed, we use the \textit{long-wavelength approximation}, where we assume that the variation in the applied field $\Ah$ is slow compared to the interatomic distance, i.e. $|\br|/\lambda \ll 1$ for some characteristic wavelength $\lambda$ of the applied field. This allows us to expand 
\begin{align}
\Ah\lel(\br_{1,2}\rer) \simeq \Ah(\bR) &\mp \frac{m_{2,1}}{M} \lel(\br\cdd\grad{\bR}\rer)\Ah(\bR) + \frac{m_{2,1}^2}{2 M} \lel(\br\cdd\grad{\bR}\rer)^2\Ah(\bR) + ... \nonumber
\end{align}
To leading order, this yields
\begin{align}
&\bL_\tm{dia} = \sum_\al q_\al \bral\times\Ah(\bral,t) \\
&\;\simeq q\bigg\{ \bR\times\lel[-\grad{\bR}\lel(\br\cdd\Ah\rer)+\br\times\Bh\rer]+ \br\times\lel[\Ah(\bR)+\frac{m_1-m_2}{M}\lel( \grad{\bR}\lel(\br\cdd\Ah\rer)-\br\times\Bh \rer)\rer]\bigg\}.\nonumber
\end{align}
We now wish to calculate the average diamagnetic angular momentum for some state of the system. As is typical, we will also assume that $m_1 \gg m_2$ to form an atom-like system and we let $m_1$ be large enough to suppress centre-of-mass motion.

To move to a quantum treatment, let us promote the coordinates to operators, such that $\bR \rightarrow \Rhh$ and $\rh \rightarrow \rhh$. There are no permanent dipoles in our model atom and so to leading-order $\avg{\rhh} = \mb{0} = \avg{\Rhh}$. The remainder is simply
\begin{align}
\avg{\Lhh_\tm{dia}} &= q\avg{\rhh\times\grad{\Rh}\lel(\rhh\cdd\Ah\rer)-\rhh\times\lel(\rhh\times\Bh\rer)}\\
&= q \intr \intR \lel|\psi(\br,\bR)\rer|^2\bigg[\br\times\grad{\bR}\lel(\br\cdd\Ah\rer) - \br\times\lel(\br\times\Bh\rer)\bigg].\nonumber
\end{align}
We further suppose that the atom is isotropically oriented. This, together with $m_1$ being large enough, allows us to express $\avg{\hat{r}_i \hat{r}_j} = \avg{\hat{r}^2}\delta_{ij}/3$ and $\lel|\psi(\br,\bR)\rer|^2 \sim |\varphi(\br)|^2\delta(\bR-\bR_A)$. The average diamagnetic angular momentum can then be expressed as
\begin{align}
\avg{\Lhh_\tm{dia}} = \frac{q \avg{\hat{r}^2}}{3}\Bh(\bR_A).
\end{align}
Finally, using the gyromagnetic ratio of $-q/2m$, we find the magnetisation
\begin{align}
\avg{\Mhh_\tm{dia}} = -\frac{q^2 \avg{\hat{r}^2}}{6m}\Bh(\bR_A),
\end{align}
per atom, which is in full agreement with, for instance, Ref.~\cite{NoltingRamakanth}[p. 88] for isotropic atoms. Thus, the general-gauge expression for $\bL_\tm{dia}$ presented here might thus have an unfamiliar form but captures the correct physics. We should note that this derivation makes no assumption about gauge nor time-dependence, and that the diamagnetic contribution appears to be instantaneous.

%\end{document}

\end{document}